\documentclass[11pt]{article}
\pdfoutput=1
\usepackage{jheppub}
\usepackage{physics}

\newcommand{\be}{\begin{equation}}
\newcommand{\ee}{\end{equation}}

\begin{document}

\preprint{}

\title{\boldmath Central charges, elliptic genera, and Bekenstein-Hawking entropy in $\mathcal{N}=(2,2)$ AdS$_3$/CFT$_2$}

\author[a]{Arash Arabi Ardehali,\footnote{a.a.ardehali@gmail.com}}
\affiliation[a]{Department of Mathematics, King's College London,\\
The Strand, London WC2R 2LS, U.K.}
\author[b]{Jiaqi Jiang,\footnote{jiaqij@princeton.edu}}
\author[b]{and Wenli Zhao\footnote{wz10@princeton.edu}}
\affiliation[b]{Joseph Henry Laboratories, Princeton University, Princeton, NJ 08544, USA}

\abstract{We consider $\mathcal{N}=(2,2)$ AdS$_3$/CFT$_2$ dualities proposed in the
large central charge limit ($c\to\infty$)
by Eberhardt. Here we propose the associated D1-D5 systems to be orbifolds of the standard $\mathcal{N}=(4,4)$ systems, thereby elevating the dualities to the finite-$c$ level on the boundary and to the quantum level in the bulk. In particular, we show that our brane systems yield low-energy sigma models whose subleading central charges match earlier predictions from bulk one-loop supergravity computations. In the case involving the Enriques surface, the finite-$c$ sigma model has a non-trivial elliptic genus which we use to microscopically explain both the Bekenstein-Hawking entropy and the subleading logarithmic correction to it for the associated macroscopic black brane.}

\maketitle \flushbottom

\section{Introduction}

AdS$_3$/CFT$_2$ is in many ways one of the simplest AdS/CFT settings  \cite{Aharony:1999ti}. The $\mathrm{SL}(2,\mathbb{Z})$ modular invariance and the infinite-dimensional Virasoro symmetry in particular give powerful handles that are not available in other dimensions. Various aspects of holography have therefore been understood in greater detail in this setting.

Remarkably however, explicit instances of AdS$_3$/CFT$_2$ with $\mathcal{N}=(2,2)$ supersymmetry (rather than with more extended SUSY) had been elusive until the last few years \cite{Datta:2017ert,Eberhardt:2017uup}.\footnote{See also \cite{Couzens:2021tnv} for an example where the CFT is implicit as the IR fixed-point of an RG flow across dimensions.} Moreover, their underlying brane configurations have been lacking until now, and as a result the new dualities have been so far checked only at the level of classical supergravity in the bulk and only in the $N\to\infty$ limit on the boundary.

Here we consider the $\mathcal{N}=(2,2)$ dualities proposed by Eberhardt in \cite{Eberhardt:2017uup}. The bulk geometry is AdS$_3\times (S^3\times M_4)/G$, with $G$ a finite group acting freely (i.e. without fixed points) on $S^3\times M_4$. The boundary CFT is a supersymmetric sigma model on $\mathrm{sym}^{\infty}(M_4/G)$. The 4-manifold $M_4$ is either $K3$ or $T^4$. In the first case $G=\mathbb{Z}_2$, such that $M_4/\mathbb{Z}_2=\mathrm{ES}$ is the \emph{Enriques surface}.\footnote{In the string theory literature the Enriques surface is sometimes referred to as $\frac{1}{2}K3$ (see e.g.~\cite{Minahan:1998vr}).} In the second case there are seven possible choices for $G$, and for each choice $M_4/G=\mathrm{HS}$ is a \emph{hyperelliptic surface}. These, of course, are essentially $G$ orbifolds of the standard $\mathcal{N}=(4,4)$ dualities involving $K3$ or $T^4$ \cite{Aharony:1999ti,Maldacena:1997re}.

By proposing associated brane configurations, we will find the boundary sigma models at finite $N$, and hence lift the dualities from the classical supergravity level to the string theory level in the bulk.

Our brane configuration consists simply of a D1-D5 system that is placed---in an appropriate manner described in Section~\ref{sec:branes}---on a background which is a quotient (by $G$) of the background of the standard $\mathcal{N}=(4,4)$ D1-D5 systems \cite{Aharony:1999ti,Maldacena:1997re}. The supergravity solutions arising this way actually have pure Ramond-Ramond flux, whereas in  \cite{Eberhardt:2017uup} the solutions have pure NS flux. The two types of solutions are related via S-duality.

We show that the leading and the subleading central charge (in the $1/N$ expansion) of the finite-$N$ sigma models that we obtain from the low-energy limit of the brane systems match respectively the bulk Brown-Henneaux central charge \cite{Brown:1986nw} and the one-loop quantum correction to it. This provides rather non-trivial evidence supporting our proposed brane configurations.

The bulk Brown-Henneaux central charge will be obtained via the standard argument involving the near-horizon limit of the brane system \cite{Maldacena:1997re,Strominger:1997eq}. The quantum correction to the Brown-Henneaux central charge, on the other hand, is extracted from the high-temperature asymptotics of the bulk one-particle superconformal index \cite{ArabiArdehali:2018mil}, also known as the one-particle NSNS elliptic genus. This procedure, which we will review in Section~\ref{sec:KKdeltaC}, was recently imported into AdS$_3$/CFT$_2$ from the AdS$_5$/CFT$_4$ context where similar prescriptions were introduced in \cite{Ardehali:2014zba,Ardehali:2014esa}. The one-particle indices that we need were computed in \cite{Eberhardt:2017uup} (and the computation will be further elucidated for the ES case in Appendix~\ref{app:1}). Their asymptotics yield  \cite{ArabiArdehali:2018mil}
\begin{equation}
\delta
c^{\mathrm{ES}}=3,\quad \text{and}\quad \delta c^{\mathrm{HS}}=0,\label{eq:deltaCintro}
\end{equation}
for the one-loop corrections to the Brown-Henneaux central charge. These numbers will serve as our entry into the realm of quantum $(2,2)$ AdS$_3$/CFT$_2$.

The finite-$N$ sigma models that we obtain from the brane setups in Section~\ref{sec:branes} allow taking a further step to study supersymmetric indices of the full CFTs---rather than just their one- or multi-trace sectors dual to the one- or multi-particle KK supergravity sectors in AdS. The full indices are expected to go beyond the KK spectra and capture various non-perturbative quantum gravitational objects in the bulk. In Section~\ref{sec:indices} we analyze the sigma model elliptic genus in the ES case to extract from it the Bekenstein-Hawking entropy of the associated BPS black brane. This is analogous to the celebrated Strominger-Vafa microstate counting \cite{Strominger:1996sh}. Here we leverage the more modern machinery developed by Sen \cite{Sen:2012cj} to also match the logarithmic correction to the entropy between the microscopic and the macroscopic side. This matching of the logarithmic correction (microscopically derived in Section~\ref{subsec:microCounting} and macroscopically reproduced in Section~\ref{subsec:macroLog}) constitutes the largest computational component of the present work.

In the HS cases, the usual CFT elliptic genus vanishes due to a target-space fermionic zero-mode. This is analogous to the situation in the standard $\mathcal{N}=(4,4)$ duality involving $T^4$, and as in that case the way forward is through a modified supersymmetric index \cite{Maldacena:1999bp}. We leave this investigation for future research.

An intriguing aspect of the brane systems discussed in this paper is a constraint on the number of D5 branes. In particular, in the ES case our sigma-model considerations in Section~\ref{sec:branes} imply that the number of D5 branes should be odd, and this oddity in turn leads to an \emph{anomalous} half-integer shift (see~Eq.~\eqref{eq:Q1shift}) in the Dirac quantization condition between the D1 and D5 brane charges in that case. We comment on this constraint from several points of view in Section~\ref{sec:open}, but more work is needed to reach a completely coherent picture of the constraint and its various avatars.

The structure of the rest of this paper is as follows. In Section~\ref{sec:KKdeltaC} we review the holographic derivation of the subleading central charges quoted in (\ref{eq:deltaCintro}). In Section~\ref{sec:branes} we present the $\mathcal{N}=(2,2)$ D-brane systems, find the finite-$N$ sigma models associated to their low-energy dynamics, and show that the leading and subleading central charges of the sigma models match the bulk expectations. In Section~\ref{sec:indices} we study the macroscopic BPS black (p-)branes that arise by exciting the D-brane systems, and in particular in the ES case give a microscopic derivation of the Bekenstein-Hawking entropy of the black brane from the elliptic genus of the finite-$N$ sigma model. We also derive a logarithmic (in the area) correction to the Bekenstein-Hawking entropy from the said elliptic genus, which we reproduce macroscopically through one-loop computations on the near-horizon geometry of the black brane. Section~\ref{sec:open} gives a brief summary of our results, and also discusses in more detail the D5-charge constraints encountered in this work. The four appendices contain technical information relied on in the main text.

\section{BPS KK spectra to lead the way}\label{sec:KKdeltaC}

In this section we review how the available BPS KK spectra on the supergravity side can be used to deduce  $\delta
c^{\mathrm{ES}}=3$ and $\delta c^{\mathrm{HS}}=0$ \cite{ArabiArdehali:2018mil}. These subleading central charges will be reproduced in the next section from the proposed brane configurations.

The starting point is the observation that at the one-loop level in the
bulk, every field in the KK supergravity spectrum on AdS$_3$
gives a holographic contribution to the subleading central charge of
the dual CFT$_2$. For a bulk field with
SL($2,\mathbb{R}$)$\times$SL($2,\mathbb{R}$) quantum numbers
$h,\bar{h}$ the contribution reads\footnote{Our notation is slightly different from the ones in \cite{Giombi:2013yva,Giombi:2013fka,Giombi:2014iua,Beccaria:2014qea}. Note in particular that (\ref{eq:deltaCforBulkField}) is a factor of $-2$ different from the result cited in Eq.~(F.3) of \cite{Beccaria:2014qea}. That is because $\delta c$ here is the contribution from a bulk field with Dirichlet boundary condition to the bulk central charge, which in \cite{Beccaria:2014qea} would be denoted by $c^+$, and as follows from the discussions in \cite{Beccaria:2014qea,Beccaria:2014xda} this is related to $c_{\text{AdS}_3}$ of \cite{Beccaria:2014qea} via $c_{\text{AdS}_3}^{\mathrm{there}}=-2c^+$. In other words $c_{\text{AdS}_3}^{\mathrm{there}}=-2\, \delta c_{\text{here}}$.} \cite{ArabiArdehali:2018mil,Giombi:2013yva,Giombi:2013fka,Giombi:2014iua,Beccaria:2014qea} (see Eq.~(3.7) in \cite{ArabiArdehali:2018mil}):
\begin{equation}
\delta
c[h,\bar{h}]=-(-1)^{2(h-\bar{h})}\frac{1}{2}(h+\bar{h}-1)((h+\bar{h}-1)^2-3(h-\bar{h})^2).\label{eq:deltaCforBulkField}
\end{equation}
When the bulk theory has $\mathcal{N}=(2,2)$ supersymmetry, the
above expression can be summed over any $(2,2)$ multiplet present in the spectrum. The
results can then be compared with the corresponding contributions of
the same multiplets to the supergravity elliptic genera
$\mathcal{I}_R(q)$, $\tilde{\mathcal{I}}_R(q)$,
$\mathcal{I}_L(\bar{q})$, and $\tilde{\mathcal{I}}_L(\bar{q})$, to
be defined shortly. At the level of
individual $\mathcal{N}=(2,2)$ multiplets, simple relations exist
between the quantities just described. Summing up those relations
over all the $(2,2)$ multiplets in the KK supergravity spectrum,
one arrives at \cite{ArabiArdehali:2018mil}
\begin{equation}
\delta c=-3\lim_{q\to1}q\partial_q(\mathcal{I}_R(q)+\tilde{\mathcal{I}}_R(q)+\mathcal{I}_L(\bar{q}=q)+\tilde{\mathcal{I}}_L(\bar{q}=q)).\label{eq:c2c2}
\end{equation}
Note in particular that long $\mathcal{N}=(2,2)$ multiplets, which do not contain BPS states, give a vanishing contribution to both $\delta c$ and the elliptic genera.

The formula (\ref{eq:c2c2}) is analogous to the relations found in
\cite{Ardehali:2014esa} (see also \cite{Ardehali:2014zba}) between the
subleading central charges $\delta a$ and $\delta c$ of
holographic 4d SCFTs and the one-particle superconformal index of their
dual supergravity theory.

We define the four types of elliptic genera appearing in (\ref{eq:c2c2}) as
\begin{equation*}
\mathcal{I}_R(q,y)=1+\mathrm{Tr}_{\mathrm{s.p.}}(-1)^{2L_0-2\bar{L}_0}q^{L_0}y^{J_0}
\bar{q}^{\bar{L}_0-\bar{J}_0},\quad
\tilde{\mathcal{I}}_R(q,y)=1+\mathrm{Tr}_{\mathrm{s.p.}}(-1)^{2L_0-2\bar{L}_0}q^{L_0}y^{J_0}
\bar{q}^{\bar{L}_0+\bar{J}_0},
\end{equation*}
\begin{equation*}
\mathcal{I}_L(\bar{q},\bar{y})=1+\mathrm{Tr}_{\mathrm{s.p.}}(-1)^{2L_0-2\bar{L}_0}q^{L_0-J_0}\bar{q}^{\bar{L}_0}\bar{y}^{\bar{J}_0},\quad
\tilde{\mathcal{I}}_L(\bar{q},\bar{y})=1+\mathrm{Tr}_{\mathrm{s.p.}}(-1)^{2L_0-2\bar{L}_0}q^{L_0+J_0}\bar{q}^{\bar{L}_0}\bar{y}^{\bar{J}_0},
\end{equation*}
specialized to $y,\bar{y}=1$. The trace is taken over the single-particle KK supergravity
Hilbert space, and $L_0,\bar{L}_0,J_0,\bar{J}_0$ are the
SL($2,\mathbb{R}$) and the U($1$)$_R$ charges\footnote{Recall that
the $\mathcal{N}=2$ superconformal algebra has a bosonic U($1$)$_R$
subgroup. We normalize the charge of this U($1$)$_R$ such that the
lowest component of a chiral multiplet in the NS sector has $J_0=L_0$.} of the
particles. In contrast with \cite{ArabiArdehali:2018mil} we have included a $+1$ vacuum contribution in the definition of the genera; this is completely inconsequential for the prescription (\ref{eq:c2c2}) which involves a derivative, and only serves to simplify the discussion by making the genera vanish in several examples. As an illustrative example, the computation of
$\mathcal{I}_R(q)$ for the ES theory is sketched in
Appendix~\ref{app:1}, and the result reads
\begin{equation}
\mathcal{I}^{\mathrm{ES}}_R(q)=\frac{1+22\sqrt{q}+13q}{1-q}.\label{eq:ESeg}
\end{equation}

For all the theories of interest to us in this paper (as well as
others treated in \cite{ArabiArdehali:2018mil}) we have $\mathcal{I}_R(q)=
\tilde{\mathcal{I}}_R(q)=\mathcal{I}_L(\bar{q}=q)=\tilde{\mathcal{I}}_L(\bar{q}=q)$. This follows from the non-chiral, CP-invariant BPS spectra of the theories.
Eq.~(\ref{eq:c2c2}) then simplifies to\footnote{The simplicity of (\ref{eq:c2c2simple}) calls for a more elegant explanation than the ``experimental'' one provided above. Such an explanation is given in \cite{ArabiArdehali:2018mil} using the notion of \emph{supersymmetric Casimir energy} \cite{Bobev:2015kza,Assel:2015nca}.}
\begin{equation}
\delta
c=-12\lim_{q\to1}q\partial_q\mathcal{I}_R(q).\label{eq:c2c2simple}
\end{equation}

Let us now apply Eq.~(\ref{eq:c2c2simple}) to the
ES theory with the index (\ref{eq:ESeg}). We get
\begin{equation}
\delta
c^{\mathrm{ES}}=-12\lim_{q\to1}q\partial_q\frac{1+22\sqrt{q}+13q}{1-q}\to\infty.\label{eq:c2c2simpleES}
\end{equation}
The divergence is not surprising. The holographic computation of the
subleading central charge involves summing
(\ref{eq:deltaCforBulkField}) over all the infinitely-many fields in
the KK supergravity spectrum, and the sum diverges unless
regularized. See \cite{Beccaria:2014qea} for a traditional
regularization in the context of AdS$_3$/CFT$_2$, and
\cite{Ardehali:2014esa,Ardehali:2014zba} for related discussions in the
AdS$_5$/CFT$_4$ context.

The regularization proposed in \cite{ArabiArdehali:2018mil} amounts to replacing
$q$ with $e^{-\epsilon}$, and expanding the right-hand side of
(\ref{eq:c2c2simple}) around $\epsilon=0$. The divergent terms are
then discarded, and the finite term is kept. With this procedure the
result of (\ref{eq:c2c2simpleES}) comes out
\begin{equation}
\delta c^{\mathrm{ES}}=-\frac{432}{\epsilon^2}+3+\dots\longrightarrow
3.\label{eq:c2c2simpleESregd}
\end{equation}

Incidentally, note that since $\mathcal{I}^{K3}_R=2\mathcal{I}^{\mathrm{ES}}_R$,
application of the prescription (\ref{eq:c2c2simple}) to the $K3$
case reproduces the correct result $\delta c=6$
\cite{Beccaria:2014qea,ArabiArdehali:2018mil}, expected from the central charge $6Q_1Q_5+6$ of the dual sigma model on
$K3^{Q_1 Q_5+1}/S(Q_1 Q_5+1)$. For the $T^4$ case the elliptic genus vanishes and therefore (\ref{eq:c2c2simple}) yields $\delta c=0$, again in
accordance with the expectation from the central charge $6Q_1Q_5$ of the dual sigma
model\footnote{We are neglecting the flat U($1$) moduli on the $T^4$
that the D5 branes wrap. Taking the U($1$) moduli into account, the
dual sigma model is on $T^4\times(T^4)^{Q_1 Q_5}/S(Q_1 Q_5)$; but
then the KK supergravity spectrum should be augmented with certain
singletons, the inclusion of which yields a non-trivial supergravity
elliptic genus, which when plugged into (\ref{eq:c2c2simple}) gives
$\delta c=6$, again matching the expectation. See
\cite{Beccaria:2014qea,ArabiArdehali:2018mil}.\label{footnote:decoupledT4index}} on
$(T^4)^{Q_1 Q_5}/S(Q_1 Q_5)$. Similarly, for the $S^3\times S^1$
case \cite{Eberhardt:2017pty} the vanishing of the bulk elliptic genus
combined with (\ref{eq:c2c2simple}) implies $\delta c=0$, once again just
as expected \cite{Eberhardt:2017pty,ArabiArdehali:2018mil}.\\

Eberhardt also discusses seven hyperelliptic manifolds
\cite{Eberhardt:2017uup}, all giving $(2,2)$ KK supergravity theories
on AdS$_3$, and all having vanishing supergravity elliptic genera.
For these, as in the $T^4$ and the $S^3\times S^1$ case, the prescription
(\ref{eq:c2c2simple}) yields
\begin{equation}
\delta c^{\mathrm{HS}}=0.\label{eq:c2c2simpleHEregd}
\end{equation}

The holographic results $\delta c^{\mathrm{ES}}=3$ and $\delta c^{\mathrm{HS}}=0$
have to be reproduced by any proposed D-brane realization of the $(2,2)$
dualities of \cite{Eberhardt:2017uup}.

\section{The brane configurations}\label{sec:branes}

In this section we propose the brane configurations underlying the $\mathcal{N}=(2,2)$ AdS$_3$/CFT$_2$ dualities of \cite{Eberhardt:2017uup} as $G$ orbifolds of the standard $\mathcal{N}=(4,4)$ D1-D5 systems. We begin by discussing the non-backreacting brane configurations, then consider backreactions, and finally present the sigma model descriptions of the low-energy brane dynamics.

\subsection{Non-backreacting D-branes}\label{subsec:orbFiber}

In this section we consider a D1-D5 system on $R\times S^1\times
R^2\times(\mathbb{C}\times M_4)/G$, with $M_4$ either $K3$ or $T^4$, and
$G$ a finite free quotient as in \cite{Eberhardt:2017uup}. For example when $M_4=K3$ we have $G=\mathbb{Z}_2$, acting via $z\to-z$ on the $\mathbb{C}$, and via the free Enriques involution on the $K3$. When $M_4=T^4$ there are seven possibilities, in the simplest of which again $G=\mathbb{Z}_2$ with the same action on the $\mathbb{C}$ while the action on the $T^4$ shifts two of the circles and reflects the other two. For the other six hyperelliptic cases $G$ and its action can be found similarly from \cite{Eberhardt:2017uup}.

We take the size of $M_4$ to be of order $\sqrt{\alpha'}$ as in \cite{Maldacena:1997re}. Then a lower-dimensional observer probing energies $\ll 1/\sqrt{\alpha'}$ perceives the geometry to be $R\times S^1\times
R^2\times(\mathbb{C}/G)$, which has an orbifold singularity at $z=0$ (because the action of $G$ on $\mathbb{C}$ is not free) as well as a deficit angle at finite $|z|$ on the $z$-plane. The orbifold singularity is of course resolved above the energy scale $1/\sqrt{\alpha'}$ once the observer begins to notice $M_4$ with $G$ acting on it freely.

As for supersymmetry, we note that the quotient does not destroy the holomorphic
three-form $\Omega^{\mathbb{C}\times M_4}$ on $\mathbb{C}\times M_4$, because the two factors in
$\Omega^{\mathbb{C}\times M_4}=\Omega^{\mathbb{C}}\times
\Omega^{M_4}$ have opposite phases---as their eigenvalues for the
quotient---compensating each other. Moreover, the first Chern class
and the Ricci curvature are not modified by the finite, free
quotient. Therefore $(\mathbb{C}\times M_4)/G$ is a non-compact CY
3-fold. It thus breaks three quarters of the supersymmetry, and we
end up with 8 supercharges, as required for the
$\mathcal{N}=(2,2)$ dualities of
\cite{Eberhardt:2017uup}.

The orbifold geometry $M_6=(\mathbb{C}\times M_4)/G$ has a divisor
with geometry $M_4/G$, which is ES when $M_4=K3$, and HS when $M_4=T^4$. This divisor can be seen as follows.
We note that $M_6$ is locally an $M_4$ bundle over $\mathbb{C}/G$,
except at the origin $z=0$ of $\mathbb{C}/G$ where the fiber should
be replaced with $M_4/G$; see the introduction section of
\cite{Sen:1996na}. The divisor corresponds to this $M_4/G$ fiber at
$z=0$. (Incidentally, the fact that the $M_4/G$ fiber can not be deformed away from
$z=0$ means that the divisor does not have moduli inside $M_6$. It is helpful to consider the toy example $(\mathbb{R}\times S^1)/\mathbb{Z}_2$, where the $\mathbb{Z}_2$ shifts the circle and reflects the line.)

Alternatively, $M_6$ is locally a $\mathbb{C}$ bundle over $M_4/G$,
with twists---corresponding to rotations in the $z$-plane---on the
$\mathbb{C}$ fiber; cf. \cite{Sen:1996na}. From this perspective, the
divisor is extended on the base, and sits on the fiber where there
is no twist, namely at $z=0$. (Again, we recommend checking the analogous statement in the toy example $(\mathbb{R}\times S^1)/\mathbb{Z}_2$.)

We consider D5 branes that are extended over
the $R\times S^1$ directions of the geometry and wrap the said divisor. Although the divisor $M_4/G$ is not spin \cite{Eberhardt:2017uup}, it is orientable as required for wrapping sources of RR flux around it, and spin$^c$ as required for vanishing Freed-Witten anomaly \cite{Freed:1999vc} of the strings ending on the D-branes wrapping it. The 10d geometry as a whole is of course spin.

We also consider D1 branes that are extended over the $R\times S^1$
directions, and made to sit at $z=0$ via a non-zero vev for $B_{\mathrm{NS}}$ (along two-cycles of $(\mathbb{C}\times M_4)/G$) that binds them to the D5 branes stuck at $z=0$; see the review \cite{David:2002wn} for an analogous discussion in the $\mathcal{N}=(4,4)$ context. We neglect this $B$-field vev in most of the discussion below though, as it is expected to be insignificant for holography of central charges and the Bekenstein-Hawking entropy, which are the topics of main focus in this work.

Note that for trivial $G$ we recover the standard $\mathcal{N}=(4,4)$ D1-D5 system \cite{Aharony:1999ti}.

\subsection*{The first encounter with the D5-charge subtleties}

We now discuss a relatively subtle aspect of the brane construction, whose significance will become more clear from a number of different perspectives in Subsection~\ref{subsec:sigma} and Section~\ref{sec:open}.

To be explicit, let us focus for the moment on the two simplest cases elaborated on at the beginning of the present section: $M_4=K3$ or $T^4$, with $G=\mathbb{Z}_2$ in both cases.

The crucial observation is that in either case pairs of D5 branes can combine to free themselves from the (torsion) cycle at $z=0$. To see this clearly, we can think of $M_6$ as a bundle over $\mathbb{C}/\mathbb{Z}_2$ as discussed above. Then the divisor that the D5 branes wrap lacks moduli precisely because the fiber is $M_4/\mathbb{Z}_2$ at $z=0$ while it is $M_4$ at $z\neq0$. However, pairs of D5 branes wrapped around $M_4/\mathbb{Z}_2$ can combine and move away from $z=0$. (It is again helpful to visualize in the toy example $(\mathbb{R}\times S^1)/\mathbb{Z}_2$, which can be thought of as an $S^1$ fiber over $\mathbb{R}/\mathbb{Z}_2$ except at the origin of $\mathbb{R}$ where the fiber is $S^1/Z_2$. Two strings wrapping the $S^1/Z_2$ fiber at the origin of $\mathbb{R}$ can combine to move away from the origin as a single string wrapping an $S^1$ fiber.)

Let us first assume that we have an even number of D5 branes. Then they can all combine in pairs, and hence are not forced to sit at $z=0$. This would add further moduli to the system that are not desirable, since we would like to have only a two-dimensional Coulomb branch---parameterized by the $R^2$ directions---per D5 brane, as appropriate for $\mathcal{N}=(2,2)$ supersymmetry. (Recall that the $(2,2)$ vector multiplet has two real scalars whereas the $(4,4)$ vector multiplet has four; see also \cite{Hanany:1997vm,Bergman:2018vqe} for other examples of $(2,2)$ brane systems and their Coulomb branches.) We are thus led to the conclusion that either with an even number of D5 branes on the orbifold background we can not realize $(2,2)$ AdS$_3$/CFT$_2$, or that somehow the branch with $z\neq0$ decouples from the branch at $z=0$ and the latter realizes the duality. Such a decoupling of different parts of the Coulomb branch would be unprecedented\footnote{There is a well-known decoupling phenomenon in the standard $(4,4)$ system \cite{Witten:1997yu}, but that is between the Higgs and the Coulomb branch, not between different parts of the same branch as needed here.} however, and in any case we will see more difficulties in Subsection~\ref{subsec:sigma} and Section~\ref{sec:open} on the way of establishing a standard AdS$_3$/CFT$_2$ correspondence with an even number of D5s.

Now let us assume an odd number of D5s.
Then, on the one hand not all the D5 branes can combine in pairs to move away from $z=0$, and on the other hand the $B$-field modulus can bind them all together forcing them to sit at $z=0$. In other words, with at least one D5 brane stuck\footnote{See \cite{Hori:1998iv,Hanany:2000fq} for early examples of branes ``stuck'' in subsets of orbifold backgrounds.} at $z=0$ for topological reasons---thereby implying an odd total number of D5 branes---we can force all the D5s to sit at $z=0$ via the $B$-field. This way the undesirable extra moduli parametrized by $z$ do not arise. We hence have a chance of realizing $(2,2)$ AdS$_3$/CFT$_2$ in the near-horizon limit of the brane system, without having to appeal to questionable decoupling assumptions as in the case with an even number of D5s.

To recap, let us denote the number of D5 branes (which coincides with the D5 charge of the system) by $Q_5$. The lesson we would like to emphasize from the preceding discussion is that there is a significant difference between the case with $Q_5$ even and that with $Q_5$ odd, and only the latter fits naturally within the standard framework of AdS$_3$/CFT$_2$. We will encounter different avatars of this subtlety in Subsection~\ref{subsec:sigma} and Section~\ref{sec:open} below.

The generalization to the other six hyperelliptic cases is that $Q_5$ should not be a multiple of $|\tilde{G}|$, where $|\tilde{G}|$ is the order of the part of $G$ that acts on $T^4$. Otherwise a non-standard decoupling argument is needed to get rid of the undesirable moduli arising from the collective motion of the combined branes along the $z$ direction.

\subsection{Backreacting D-branes and p-branes}\label{subsec:pBrane}

Since our D1-D5 configurations are essentially orbifolds of the
original $\mathcal{N}=(4,4)$ D1-D5 systems on $K3$ or $T^4$, we expect them to
lead to p-brane solutions in IIB supergravity that are simply
orbifolds of the p-brane solutions of the original $\mathcal{N}=(4,4)$
systems. The decoupling-limit argument of
\cite{Maldacena:1997re} would then realize the $(2,2)$ dualities of
\cite{Eberhardt:2017uup} in string theory.\\

To be more precise, we write down the explicit metric, dilaton, and
3-form flux of the p-brane solution in IIB supergravity\footnote{A puzzling aspect of this solution is that it is homogeneous in the $M_4$ directions, while the D1 branes are not necessarily uniformly distributed along $M_4$. See \cite{Surya:1998dx,Marolf:1999uq} for an explanation of this point.} (see e.g. \cite{Dhar:1999ax})
\begin{equation}
\begin{split}
e^{-2\phi}&=f_5/f_1,\\
\mathrm{d}s^2&=f_1^{-1/2}f_5^{-1/2}\mathrm{d}x_{||}^2+f_1^{1/2}f_5^{1/2}(\mathrm{d}r^2+r^2\mathrm{d}\Omega_3^2)+f_1^{1/2}f_5^{-1/2}\mathrm{d}x_{M_4}^2,\\
F_3&=2r_5^2\epsilon_3+2r_1^2 e^{+2\phi}\ast_{10}\epsilon_7,\\
f_i&:=1+r_i^2/r^2\quad\ i=1,5,\label{eq:pBrane}
\end{split}
\end{equation}
where $\mathrm{d}x_{||}^2=-\mathrm{d}t^2+\mathrm{d}x^2$, with $x$
the coordinate along the D1-branes. The radial
coordinate on $R^2\times \mathbb{C}$ is parameterized by $r$. The forms $\epsilon_3$ and
$\epsilon_7$ are the volume forms of a three-cycle $\mathcal{C}_3$ and the
seven-cycle $\mathcal{C}_7$ at $r=1$ inside $R^2\times( \mathbb{C}\times M_4)/G$.
We can more explicitly describe $\mathcal{C}_3$ as the three-cycle descending from the unite-radius $S^3\subset R^2\times\mathbb{C}$ in
$(S^3\times M_4)/G\subset R^2\times( \mathbb{C}\times M_4)/G$ (see the remarks below (\ref{eq:volFormulas})), and $\mathcal{C}_7$ as $(S^3\times M_4)/G$ itself.

Importantly, as long as we do not use \emph{global} relations to
relate the parameters $r_i$ to the D-brane charges, the solution in
(\ref{eq:pBrane}) is \emph{locally} exactly the same as that of the
standard D1-D5 p-brane, also known as the 6d black string. The IIB
supergravity equations are hence obviously satisfied. The global
differences are: $i)$ that we have assumed that the ranges of
various coordinates are related to those in the standard 6d black
string via appropriate identifications due to $G$, and
$ii)$ that we have written $F_3$ in a form language more suitable to
the (ten-dimensional) orbifold geometry.\footnote{In the literature sometimes a 6d Hodge star is used in writing similar three-form fluxes, which is not suitable for the orbifolded geometries of our interest here. See for instance \cite{Dabholkar:1997rk}. Note that while the 6d Hodge star in Eq.~(1.8) of \cite{Dabholkar:1997rk} is multiplied by $e^{-2\phi}$, the 10d Hodge star should be multiplied by $e^{+2\phi}$ as in (\ref{eq:pBrane}) so it cancels extra factors arising from $\sqrt{\mathrm{det}g^{(10d)}}$.}

Once we integrate $F_3$ at $r\to\infty$ to relate the local parameters $r_i$ to the global charges $Q_i$ of
the D-branes, we find (for the pre-factors, compare with \cite{Polchinski:1995mt})
\begin{equation}
Q_5=\frac{1}{4\pi^2\alpha'
g}\int_{\mathcal{C}_3}F_3=\frac{r_5^2}{\alpha' g},\quad
Q_1=\frac{1}{(4\pi^2\alpha')^3
g}\int_{\mathcal{C}_7}\ast_{10}F_3=\frac{r_1^2 v}{\alpha'
g}.\label{eq:fluxIntegration}
\end{equation}
Therefore
\begin{equation}
r_5^2=g\alpha' Q_5,\quad\ r_1^2=\frac{g\alpha'
Q_1}{v},\label{eq:r1andr5}
\end{equation}
just like the standard D1-D5 systems, \emph{but now with}
\begin{equation}
v:=\frac{\mathrm{vol}(M_4)/((2\pi)^4 \alpha'^2)}{|G|},\label{eq:v}
\end{equation}
where $|G|$ is the order of the part of $G$ that acts on
$\mathbb{C}$ (or equivalently, on the $S^3$ at $r=1$ in $R^2\times
\mathbb{C}$). In the ES case for example $|G|=2$. To evaluate the integrals in
(\ref{eq:fluxIntegration}) we have used
\begin{equation}
\int
\epsilon_7=\frac{\mathrm{vol}(S^3)\times\mathrm{vol}(M_4)}{|G|},\quad
\mathrm{\ and}\quad\ \int \epsilon_3=\mathrm{vol}(S^3).\label{eq:volFormulas}
\end{equation}
Note that the difference with the standard D1-D5 systems originates
from the integral of $\epsilon_7$ now being smaller due to $G$. To
arrive at the first integral above, we can use the fact that $(S^3
\times M_4)/G$ is an $M_4$ fiber over $S^3/G$ except on a set of
measure zero on the base where $G$ leaves a circle of $S^3$ fixed.
Therefore $\mathrm{vol}((S^3 \times
M_4)/G)=\mathrm{vol}(S^3/G)\times\mathrm{vol}(M_4)$, yielding the
desired result. For the second integral we note that $(S^3 \times
M_4)/G$ is an $S^3$ fiber over $M_4/G$, with twists on the $S^3$ as
one goes around the cycles in $M_4/G$ that are
introduced by the $G$-quotient. To compute the integral of $\epsilon_3$ we can sit at a
point on $M_4/G$, and so we need not notice the topological
complications that arise upon traversing the cycles of $M_4/G$; we end
up with the volume of the fiber ($\mathrm{vol}(S^3)$) as claimed.\\

As a byproduct of the above discussion we obtain the value of the
3d Newton's constant in the AdS$_3$ space arising from the
decoupling limit of the p-brane solution (\ref{eq:pBrane}). In the
near-horizon limit the geometry becomes \cite{Maldacena:1997re}
\begin{equation}
\mathrm{d}s_{10}^2=\alpha'(\mathrm{d}s_{\mathrm{AdS}_3}^2+\ell^2
\mathrm{d}\Omega_3^2+\sqrt{\frac{v Q_1}{Q_5}}\mathrm{d}x_{M_4}^2),
\end{equation}
where
\begin{equation}
\mathrm{d}s_{\mathrm{AdS}_3}^2=\frac{U^2}{\ell^2}\mathrm{d}x_{||}^2+\frac{\ell^2}{U^2}\mathrm{d}U^2,\quad
U=\frac{r}{\alpha'},\quad \ell=\left(\frac{g^2 Q_1
Q_5}{v}\right)^{1/4}. \label{eq:nhGeometry}
\end{equation}
Therefore
\begin{equation}
G^N_{\mathrm{AdS_3}}=\frac{G^N_{10}}{\mathrm{vol}((S^3 \times
M_4)/G)}=\frac{8\pi^6 g^2 \alpha'^4}{2\pi^2 R_{S^3}^3\times (2\pi)^4
\alpha'^2 v}=\frac{\sqrt{\alpha'}(g^2 Q_1 Q_5/v)^{1/4}}{4Q_1
Q_5},\label{eq:GAdS3}
\end{equation}
where we used $R_{S^3}=\sqrt{\alpha'}\ell$.

The important conclusion is that, from (\ref{eq:nhGeometry}) and
(\ref{eq:GAdS3}), the Brown-Henneaux central charge comes out
\begin{equation}
c_{0}=3R_{\mathrm{AdS}_3}/2G^N_{\mathrm{AdS_3}}=6Q_1
Q_5,\label{eq:BHcc}
\end{equation}
where we used
$R_{\mathrm{AdS}_3}(=R_{S^3})=\sqrt{\alpha'}\ell$.

Note that the quotient by $G$ does not affect the expression of $c_{0}$ in terms of $Q_1,Q_5$,
and in particular the expression is the same as that of the standard
D1-D5 systems corresponding to trivial $G$.

We will see below that the sigma model describing the low-energy
D-brane dynamics reproduces precisely this central charge.

\subsection{Sigma model description of the low-energy D-brane
dynamics}\label{subsec:sigma}

Before investigating the low-energy sigma models of the brane setups just described, we review the analogous aspects of the standard $(4,4)$ D1-D5 systems.

\subsubsection*{Recap of the standard $(4,4)$ story}

The standard $(4,4)$ constructions correspond to trivial $G$, with the
branes on the background $S^1\times R\times
R^2\times\mathbb{C}\times M_4$.

The IR limit of the brane system is described by a superconformal
sigma model with target-space dimension (we consider SU($N$)
instantons, and do not count flat U($1$) moduli)
\cite{Vafa:1995bm,Atiyah:1978wi,Bershadsky:1995qy}
\begin{equation}
\mathrm{dim}\mathcal{M}^{Q_5}_{k_1}=4k_1Q_5-(Q_5^2-1)(\frac{\chi+\tau}{2}),\label{eq:dimNandk}
\end{equation}
where $\chi$ and $\tau$ are the Euler characteristic and the
signature of $M_4$ respectively.\footnote{The formula in
\cite{Atiyah:1978wi} has $\frac{\chi-\tau}{2}$ instead of
$\frac{\chi+\tau}{2}$. That is because \cite{Atiyah:1978wi} considers
self-dual 4-manifolds, while we focus on anti-self-dual 4-manifolds. A change of orientation with
$\tau\to-\tau$ relates the two.} Here $Q_5$ stands for the D5-brane charge of the
system, which coincides with the number of D5 branes. On the other hand, the number of
D1 branes, denoted by $k_1$, does \emph{not} necessarily
coincide with the D1-brane charge of the system for the following reason.

One of the most beautiful aspects of the D1-D5 system is that due to
an \emph{anomalous I-brane inflow} of D1 charge from the D5 branes, the D1-brane charge of the system becomes~\cite{Green:1996dd}
(see also \cite{Bershadsky:1995qy,Cheung:1997az,Minasian:1997mm} for related work, and Section~4 of \cite{Harvey:2005it} for a review of the relevant material)
\begin{equation}
Q_1=k_1+(\tau/16)Q_5.\label{eq:Q1}
\end{equation}
Note that we have used the Hirzebruch signature theorem to write the right-hand side
in terms of $\tau$ instead of the more commonly used Pontryagin
class (cf. \cite{Bershadsky:1995qy,Vafa:1995bm}).

Since we are dealing with a supersymmetric sigma model, each
boson (with $c=1$) is accompanied by a fermion (with
$c=\frac{1}{2}$), and thus
$c=\frac{3}{2}\mathrm{dim}\mathcal{M}^{Q_5}_{k_1}$. This, together with
(\ref{eq:dimNandk}) and (\ref{eq:Q1}), yields
\begin{equation}
c^{K3}=6Q_1Q_5+6,\label{eq:cK3}
\end{equation}
and
\begin{equation}
c^{T^4}=6Q_1Q_5.\label{eq:cT4}
\end{equation}
We have used $\chi(K3)=24$, $\tau(K3)=-16$, and
$\chi(T^4)=\tau(T^4)=0$.

Assuming that the sigma-model target is of the form $\mathrm{sym}^{N}(M_4)$, from the above central charges and the fact that for a single copy of $K3$ or $T^4$ the central charge is $6$, we infer that the target spaces are
$\mathrm{sym}^{Q_1 Q_5+1}(K3)$ and $\mathrm{sym}^{Q_1 Q_5}(T^4)$
respectively \cite{Vafa:1995bm}.

\subsubsection*{The $(2,2)$ construction}

We now consider the $(2,2)$ systems arising for
non-trivial $G$.

Since the D5 branes now wrap a divisor of the form ES or HS, using
(\ref{eq:dimNandk}) and (\ref{eq:Q1}), together with $\chi(\mathrm{ES})=12$, $\tau(\mathrm{ES})=-8$,
$\chi(\mathrm{HS})=\tau(\mathrm{HS})=0$, we find the
corresponding central charges to be
\begin{equation}
c^{\mathrm{ES}}=6Q_1Q_5+3,\label{eq:cES}
\end{equation}
\begin{equation}
c^{\mathrm{HS}}=6Q_1Q_5.\label{eq:cHS}
\end{equation}

The leading $6Q_1Q_5$ pieces in (\ref{eq:cES}) and (\ref{eq:cHS}) match the
Brown-Henneaux central charge in (\ref{eq:BHcc}). Recall that this is the same expression in terms of $Q_1,Q_5$ as that in the standard $(4,4)$
cases.

More interestingly, the subleading pieces in (\ref{eq:cES}) and (\ref{eq:cHS}) match
the one-loop corrections to the Brown-Henneaux central charge, as
computed from the BPS KK spectra in Section~\ref{sec:KKdeltaC}.
This is one of the main results of the present paper.\\

The results (\ref{eq:cES}) and (\ref{eq:cHS}), together with the fact that a single copy of ES or HS has central charge $\frac{3}{2}\times 4=6$, can now guide us in finding the finite-$N$
counterparts of Eberhardt's supersymmetric sigma models on
$\mathrm{sym}^\infty (\mathrm{ES})$ and $\mathrm{sym}^\infty (\mathrm{HS})$.

In the ES case the natural finite-$N$ candidate with
central charge $6Q_1Q_5+3$ is the supersymmetric sigma model with
target space
\begin{equation}
  \mathrm{sym}^{Q_1Q_5+\frac{1}{2}} (\mathrm{ES}).\label{eq:ESsigma}
\end{equation}
Note that this makes sense only if $Q_1
Q_5+\frac{1}{2}$ is an integer. Since in the ES case (see (\ref{eq:Q1}))
\begin{equation}
Q_1=k_1-Q_5/2,\label{eq:Q1shift}
\end{equation} 
the requirement that $Q_1
Q_5+\frac{1}{2}$ be an integer translates to the constraint that 
$Q_5$ be odd. This is one manifestation of the odd-$Q_5$ constraint mentioned in Subsection~\ref{subsec:orbFiber} above. We will discuss another manifestation of it in Section~\ref{sec:open}.

In the HS cases the natural supersymmetric sigma model candidate is
that with target space
\begin{equation}
\mathrm{sym}^{Q_1Q_5} (\mathrm{HS}).\label{eq:HSsigma}
\end{equation}
This is a well-defined space for any integer
$Q_5$. So the constraint mentioned in Subsection~\ref{subsec:orbFiber} is not recognizable at the level of the sigma-model target space in the HS cases.

\section{BPS state counting beyond KK spectra}\label{sec:indices}

The finite-$N$ sigma models that we obtained in the last section allow taking a further step to compute supersymmetric indices of the full CFTs. The full indices encode BPS states beyond the KK supergravity spectra.

In the present section we study the CFT elliptic genus in the ES case and extract from it asymptotic degeneracies associated to black branes. The black branes in question are of course simply $G$ orbifolds of the famous black branes studied by Strominger and Vafa \cite{Strominger:1996sh}.

\subsection{The black branes and their Bekenstein-Hawking entropy}

The black branes arise by adding $n$ units of left-moving momentum on the $S^1$. This excited system is referred to as the D1-D5-P system. The momentum modifies the p-brane geometry (\ref{eq:pBrane}) to (see e.g. \cite{Hyun:1997jv})
\begin{equation}
\begin{split}
\mathrm{d}s^2=f_1^{-1/2}f_5^{-1/2}\big(-\mathrm{d}t^2+\mathrm{d}x^2+(f_n-1)(\mathrm{d}t-\mathrm{d}x)^2\big)+f_1^{1/2}f_5^{1/2}(\mathrm{d}r^2+r^2\mathrm{d}\Omega_3^2)+f_1^{1/2}f_5^{-1/2}\mathrm{d}x_{M_4}^2,\label{eq:pBraneExcited}
\end{split}
\end{equation}
with 
\begin{equation}
f_i=1+r_i^2/r^2,\ i=1,5,\qquad f_n:=1+r_n^2/r^2,\quad r_n^2=\frac{g^2\alpha'^2
n}{vR_{S^1}^2},\label{eq:rn}
\end{equation}
where $R_{S^1}$ is the radius of the circle the D1-branes are wrapped on (see \cite{David:2002wn} for more details and references). The black brane geometry is again locally exactly the same as the Strominger-Vafa black brane. The global differences are that the ranges of the parameters are now different due to identifications by $G$, and that the relations between $r^2_1$ and $Q_1$, and between $r^2_n$ and $n$, are different due to $v$ being smaller now by a factor of $|G|$ as in (\ref{eq:v}).

Note that while the Strominger-Vafa black brane can be reduced on $S^1\times M_4$ to give a lower-dimensional picture as a 5d black hole, in our case since $G$ mixes $M_4$ with the other parts of the geometry the only \emph{smooth} lower-dimensional picture is that obtained by reducing on the $S^1$, which is a 9d black brane. A low-energy observer probing energies $\ll \frac{1}{R_{S^1}},\frac{1}{\sqrt{\alpha'}}$ would of course perceive a 5d geometry in our case too, but with an \emph{orbifold singularity}, as well as a deficit angle on a plane at large $r$; see the comments at the beginning of the previous section. In Section~\ref{subsec:macroLog} below, we compute certain quantum effects both in a way that is more natural from the (singular) 5d perspective of the low-energy observer, and in a way more natural from the (smooth) 9d point of view, finding agreement.

The entropy of the 9d Strominger-Vafa black brane is given by the Bekenstein-Hawking formula (cf. \cite{David:2002wn})
\begin{equation}
\begin{split}
    S_{\mathrm{SV}}=\frac{A}{4G_9}=\frac{(2\pi^2 R_h^3)\mathrm{vol}(M_4)}{4G_9}&=\frac{(2\pi^2 r_1r_5r_n)\mathrm{vol}(M_4)}{4G_9}\\
    &=2\pi\sqrt{Q_1Q_5n},\label{eq:Ssv}
\end{split}
\end{equation}
where $G_9=G_{10}/2\pi R_{S^1}=8\pi^6 g^2\alpha'^4/2\pi R_{S^1}$ is the 9d Newton constant.

For our 9d black branes the expression for the entropy in terms of the metric parameters $r_{1,5,n}$ is smaller by a factor of $|G|$ due to the orbifold identifications reducing the horizon area. However, according to (\ref{eq:r1andr5}) and (\ref{eq:rn}), when writing $r_1$ and $r_n$ in terms of the global charges $Q_1$ and $n$, we have factors of $\sqrt{|G|}$ arising from the denominator of $v$ in (\ref{eq:v}). Hence, in our case (\ref{eq:Ssv}) is modified to
\begin{equation}
\begin{split}
    S_{\mathrm{SV}/G}&=\frac{\left((2\pi^2 r_1r_5r_n)\mathrm{vol}(M_4)\right)/|G|}{4G_9}\\
    &=\frac{2\pi\sqrt{(|G|Q_1)Q_5(|G|n)}}{|G|}=2\pi\sqrt{Q_1Q_5n}.\label{eq:S2c2}
\end{split}
\end{equation}
In other words the expression for the entropy in terms of the charges $Q_1,Q_5,n$ is exactly the same as that in the standard (4,4) case of Strominger-Vafa (where $G$ was trivial)! This should have been expected in fact: general AdS$_3$/CFT$_2$ considerations imply that the entropy of the black branes is reproduced by the Cardy formula, which in turn is fixed by the central charge; on the other hand, we had found in (\ref{eq:BHcc}) that the leading central charge of the $(2,2)$ cases has the same expression in terms of $Q_{1,5}$ that it has in the standard $(4,4)$ cases. The puzzle should have been how validity of the same Cardy formula (with the same leading-order central charge) for both trivial and nontrivial $G$ is consistent with the fact that non-trivial $G$ reduces the horizon area; the resolution, clear from the above discussion, lies in the relation between the ``local'' parameters $r_{1,n}$ and the ``global'' charges $Q_1,n$.

To be clear, general AdS$_3$/CFT$_2$ considerations together with the Cardy formula and the Brown-Henneaux central charge (\ref{eq:BHcc}) do account microscopically for the Bekenstein-Hawking entropy (\ref{eq:S2c2}) in the $(2,2)$ cases as well. What we investigate below is the more non-trivial question of whether the macroscopic entropy (\ref{eq:S2c2}) can be accounted for also by appropriate supersymmetric indices of the microscopic CFTs. The answer is positive in the ES case. We leave the analogous investigation of the HS cases (requiring modified supersymmetric indices as in \cite{Maldacena:1999bp}) to future work.

Also, the Bekenstein-Hawking entropy is corrected quantum mechanically by a logarithmic term (in the area) that can not be captured by universal Cardy-like formulas or explained by general AdS/CFT considerations. It thus calls for a direct microscopic calculation, which we perform in the ES case with the aid of the CFT elliptic genus. The matching of this logarithmic piece with the macroscopic result derived below provides a more refined check of the duality.

\subsection{BPS microstate counting in the ES case}\label{subsec:ESindex}

Here we imitate Sen's discussion
\cite{Sen:2012cj} on the standard $(4,4)$ duality involving $K3$.

\subsubsection{The seed elliptic genus}

Our seed CFT target space being ES$=K3/\mathbb{Z}_2$ instead of $K3$, the seed elliptic genus becomes half as much as that of $K3$ \cite{Eberhardt:2017uup}:
\begin{equation}
    \chi(\mathrm{ES})=4\left(\frac{\theta_2(z,\tau)^2}{\theta_2(\tau)^2}+\frac{\theta_3(z,\tau)^2}{\theta_3(\tau)^2}+\frac{\theta_4(z,\tau)^2}{\theta_4(\tau)^2}\right).\label{eq:ES1EG}
\end{equation}
A quick way to derive this result would be through Eq.~(2.5) of \cite{Manschot:2007ha}, using $\chi(\mathrm{ES})=12$, $\tau(\mathrm{ES})=-8$.

Note that (\ref{eq:ES1EG}) is the RR elliptic genus. The relation with the NSNS elliptic genera of the type encountered in Section~\ref{sec:KKdeltaC} is through spectral flow, as discussed explicitly in \cite{Eberhardt:2017uup}.\footnote{More precisely, the plethystic logarithm of the large-$N$ limit of $\chi(\mathrm{sym}^N(\mathrm{ES}))$ in the ``confined'' phase, spectrally flowed to the NSNS sector, is the bulk one-particle elliptic genus discussed in Section~\ref{sec:KKdeltaC} (modulo a +1 vacuum contribution depending on the definition) \cite{Eberhardt:2017uup}. In this section we are going to study the ``deconfined'' phase of $\chi(\mathrm{sym}^N(\mathrm{ES}))$ to make contact with the macroscopic black brane. In the Cardy limit ($|\tau|\to0$) the control-parameter triggering the deconfinement transition is $z$.}

\subsubsection{Generating function of the symmetric orbifold}

The elliptic genus being half that of $K3$, the corresponding
generating function
\begin{equation}
    \mathcal{Z}:=\sum_{N=0}^\infty p^N \chi(\mathrm{sym}^N(\mathrm{ES})),
\end{equation}
becomes the square root of $\mathcal{Z}_{\mathrm{DMVV}}$ \cite{Dijkgraaf:1996xw}. As a
result, the corresponding Siegel modular form is $\sqrt{\Phi_{10}}$, or in other words
\begin{equation}
    \mathcal{Z}(p,q,y)=\frac{1}{\sqrt{\Phi_{10}(\rho,\tau,z)}},\label{eq:ESsiegel}
\end{equation}
where $p=e^{2\pi i\rho},q=e^{2\pi i\tau},y=e^{2\pi i z}$.

\subsubsection{Microscopic black brane degeneracy (leading order match)}\label{subsec:microCounting}

In our case $N=c/6=Q_1Q_5$. The degeneracies are hence derived from the coefficient of $p^N=e^{2\pi i\rho Q_1Q_5}$ in $\mathcal{Z}$. Moreover, we want to extract the coefficient of $q^n=e^{2\pi i\tau n}$ and $y^J=e^{2\pi i z J}$. This is accomplished simply by extracting the appropriate residues of (\ref{eq:ESsiegel}) via
\begin{equation}
\tilde{d}_{\mathrm{micro}}^{\mathrm{ES}}(n,Q_1,Q_5,J)\simeq\oint \frac{\mathrm{d}q}{2\pi i q}\oint
\frac{\mathrm{d}p}{2\pi i p}\oint\frac{\mathrm{d}y}{2\pi i y}\  e^{-2\pi i(\tau n+\rho
Q_1Q_5+Jz)}\frac{1}{\sqrt{\Phi_{10}(\rho,\tau,z)}},\label{eq:deg&chiEStilde}
\end{equation}
with small enough contours around the origin. For simplicity we set the angular-momentum quantum number $J$ to zero; otherwise we would have to deal with an orbifolded BMPV \cite{Breckenridge:1996is} geometry rather than an orbifolded Strominger-Vafa geometry. We also write the integrals in terms of $\tau,\rho,z$. We end up with\footnote{This equation is precisely the analog of Sen's Eq.~(5.2) in \cite{Sen:2012cj}. Compared to that equation we have suppressed a power of $(e^{i\pi z}-e^{-i\pi z})$ in the integrand (due to fermion zero-modes and the center-of-mass motion of the D1-D5 system), as it would not contribute either to the Bekenstein-Hawking entropy or to its logarithmic correction.}
\begin{equation}
\tilde{d}_{\mathrm{micro}}^{\mathrm{ES}}(n,Q_1,Q_5,0)\simeq\int \mathrm{d}\tau\int
\mathrm{d}\rho\int\mathrm{d}z\  e^{-2\pi i(\tau n+\rho
Q_1Q_5)}\frac{1}{\sqrt{\Phi_{10}(\rho,\tau,z)}}.\label{eq:deg&chiEStilde2}
\end{equation}

The remaining saddle-point analysis is by now standard
(cf. \cite{Sen:2012cj,Belin:2016knb}). We can write (\ref{eq:deg&chiEStilde2}) using modular properties of $\Phi_{10}$ as
\begin{equation}
\tilde{d}_{\mathrm{micro}}^{\mathrm{ES}}(n,Q_1,Q_5,0)\simeq\int \mathrm{d}\hat{\tau}\int
\mathrm{d}\hat{\rho}\int\mathrm{d}\hat{z}\ e^{-2\pi i(\tau n+\rho
Q_1Q_5)}\mathrm{det}(C\Omega+D)^8\frac{1}{\sqrt{\Phi_{10}(\hat{\Omega})}},\label{eq:deg&chiESmod}
\end{equation}
where $\hat{\Omega}$ stands for $\hat{\rho},\hat{\tau},\hat{z}$,
with
\begin{equation}
\tau=\frac{1}{2\hat{z}-\hat{\rho}-\hat{\tau}},\quad
\rho=\frac{\hat{z}^2-\hat{\rho}\hat{\tau}}{2\hat{z}-\hat{\rho}-\hat{\tau}},\quad
z=\frac{\hat{z}-\hat{\rho}}{2\hat{z}-\hat{\rho}-\hat{\tau}},\label{eq:omegaHat}
\end{equation}
so that
$\mathrm{det}(C\Omega+D)=(2\hat{z}-\hat{\rho}-\hat{\tau})^{-1}$.
Using the fact that near $\hat{z}=0$ we have
\begin{equation}
\frac{1}{\sqrt{\Phi_{10}(\hat{\Omega})}}=\frac{1}{2\pi
i\hat{z}}\eta(\hat{\tau})^{-12}\eta(\hat{\rho})^{-12}+\dots,\label{eq:chiHatExp}
\end{equation}
we can perform the contour integral by picking up the residue of the
simple pole of (\ref{eq:deg&chiESmod}) at $\hat{z}=0$, and end up
with
\begin{equation}
\tilde{d}_{\mathrm{micro}}^{\mathrm{ES}}(n,Q_1,Q_5,0)\simeq\int \mathrm{d}\hat{\tau}\int
\mathrm{d}\hat{\rho}\  e^{-\frac{2\pi
i}{\hat{\rho}+\hat{\tau}}(Q_1Q_5\hat{\tau}\hat{\rho}-n)}\eta(\hat{\tau})^{-12}\eta(\hat{\rho})^{-12}(\hat{\rho}+\hat{\tau})^{-8}.\label{eq:deg&res}
\end{equation}
After introducing $\tau_{1,2}$ through $\hat{\rho}=\tau_1+i\tau_2$
and $\hat{\tau}=-\tau_1+i\tau_2$, our expression simplifies to
\begin{equation}
\tilde{d}_{\mathrm{micro}}^{\mathrm{ES}}(n,Q_1,Q_5,0)\simeq\int \mathrm{d}\tau_1\int
\mathrm{d}\tau_2\ e^{\frac{\pi
}{\tau_2}(Q_1Q_5(\tau_1^2+\tau_2^2)+n)}\eta(-\tau_1+i\tau_2)^{-12}\eta(\tau_1+i\tau_2)^{-12}\tau_2^{-8}.\label{eq:deg&taus}
\end{equation}
Scaling $n\sim Q_{1,5}\sim\Lambda\to\infty$, we get the saddle point from the
exponential, lying at $\tau_1=0$ and $\tau_2=\sqrt{n/Q_1Q_5}$. From
computing the square root of the second derivative of the
exponential with respect to $\tau_{1,2}$ we find the effective width of the
$\tau_{1,2}$ integrals to be both $\Lambda^{-5/4}$.
Finally, since the integrals can be replaced with their effective width, we
get two factors of $\Lambda^{-5/4}$ from the two integrals, which
together with $\eta(i\tau_2)\sim \tau_2^{-1/2}$, and
$\tau_2\sim\Lambda^{-1/2}$, give all in all
\begin{equation}
\tilde{d}_{\mathrm{micro}}^{\mathrm{ES}}(n,Q_1,Q_5,0)\simeq
e^{2\pi\sqrt{Q_1Q_5n}}\Lambda^{-9/2}.\label{eq:degFinal}
\end{equation}
In other words, the leading entropy matches (\ref{eq:S2c2})!

The log correction to the entropy comes out from (\ref{eq:degFinal}) to be
\begin{equation}
-9\log\Lambda^{1/2}.\label{eq:logCor}
\end{equation}
Before reproducing the log correction (\ref{eq:logCor}) from a macroscopic computation, it is worth noting that in \cite{Sen:2012cj} a different microscopic ensemble with SU(2) quantum number $\vec{J}^2$ fixed to zero, and degeneracies denoted $d$ rather than $\tilde{d}$, was considered for the non-rotating black holes. This is because according to Sen's entropy function formalism the choice of ensemble on the microscopic side is dictated by the symmetries preserved in the near-horizon geometry on the macroscopic side. For our black brane, in the near-horizon geometry AdS$_2\times (S^3\times K3)/\mathbb{Z}_2\times S^1$, the $Z_2$ orbifold breaks the left-handed SU(2) down to U(1) \cite{Eberhardt:2017uup}. Therefore the issue of fixing the SU(2) quantum number $\vec{J}^2$ does not arise for us, and the degeneracies $\tilde{d}$ rather than $d$ are relevant, even for $J=0$.

We now explain how this logarithmic correction can be reproduced macroscopically through a one-loop supergravity computation.

\subsubsection{Macroscopic computation of the log correction (subleading order match)}\label{subsec:macroLog}

\subsubsection*{The set up}
Computing the logarithmic correction to the black hole entropy has been of major interest for a long time, as it often captures a leading quantum correction to the Bekenstein-Hawking formula (see e.g.~\cite{Sen:2011ba,Keeler:2014bra} and references therein). Moreover, in a limit where the black hole is large, it exhibits features that allow its determination completely within one-loop supergravity (regardless of possible $\alpha'$ or higher-genus corrections; see \cite{Sen:2012dw,Bhattacharyya:2012ye}). More precisely, for asymptotically (macroscopically, locally) flat black holes at zero temperature, if one scales the mass $M$ and charges $Q_i$ of a black hole to be large while keeping the black hole extremal,
\begin{equation}
    Q_i\sim \Lambda,\quad M \sim \Lambda,\qquad T=0, 
\end{equation}
the logarithmic correction to the black hole entropy can be computed from the gravitational partition function on the near horizon geometry AdS$_2\times M_{d-2}$, where $M_{d-2}$ is the event horizon (possibly containing some internal compact factor that does not grow as $\Lambda\to\infty$). The Strominger-Vafa black hole corresponds to $M_8=S^3\times K3\times S^1$, with the $S^3$ growing as $\Lambda\to\infty$. Note that from an AdS$_3$/CFT$_2$ viewpoint, the AdS$_2\times S^1$ part of the geometry is a BTZ black hole for large $n$ (the D1-D5-P system) and an empty AdS$_3$ for $n=0$ (the D1-D5 system). 

The scaling limit not only guarantees the validity of the semi-classical approximation, but also is computationally convenient as it allows focusing on a single logarithmic term (i.e. $\log\Lambda$) instead of having to consider several such terms (e.g. $\log M$, $\log Q_i$, etc). As a further simplification, when the number of spacetime dimensions that grow as $\Lambda\to\infty$ is odd, the logarithmic correction is determined entirely by the zero modes (see e.g. \cite{Bhattacharyya:2012ye}). The AdS$_2$ part of the geometry in particular only supports three types of zero modes: gravitons, gravitinos, and one forms \cite{CAMPORESI199457}.

Below we reduce the IIB theory either on $S^1$ and perform a 9-dimensional gravitational path integral on $AdS_2\times\left(S^3\times K3\right)/\mathbb{Z}_2$, or adopt a low energy ($E\ll \frac{1}{R_{S^1}},\frac{1}{\sqrt{\alpha'}}$) perspective and carry out a 5-dimensional path integral on $AdS_2\times S^3/\mathbb{Z}_2.$ Since the logarithmic correction to the entropy is expected to be accessible in the infrared \cite{Sen:2011ba}, and since in both approaches we are keeping the AdS$_2$ and $S^3$ pieces of the geometry which are the parts that grow as $\Lambda\to\infty$, we assume that reduction on the $S^1$ or adopting the low-energy perspective does not invalidate our calculation of the logarithmic term.

However, both approaches encounter new subtleties that prevent a direct application of the existing formalism outlined in \cite{Bhattacharyya:2012ye,Liu:2017vll}. For the 9d calculation, under the scaling limit, the size of $K3$ is fixed while the size of AdS$_2$ and $S^3$ becomes large. The additional dimensionful parameter, the size of $K3$, introduces a new subtlety in the argument of \cite{Bhattacharyya:2012ye}. We shall show that even though there are two distinct scales, the logarithmic contribution is determined completely by zero modes. Such a conclusion, in fact, only relies on the number of `large' dimensions under the scaling limit to be odd. For the 5d calculation, the orbifold singularity appears to present a difficulty at first. However, it has been well-understood that once twisted-sector states are added to backgrounds with orbifold singularity, quantum corrections should match with the microscopic computations. See in particular \cite{GUKOV199823,Gadde:2009dj, Ardehali:2013xya}. In our case, the twisted sector states are massive (with masses of order $1/\sqrt{\alpha'}$ as discussed below) because the orbifold singularity is actually resolved at the $K3$ scale. Therefore they do not contribute to the logarithmic term, and hence the untwisted sector on the orbifold background should suffice for matching with the microscopic side. We will consider both calculations independently below, and show that they reproduce (\ref{eq:logCor}).

\subsubsection*{9d calculation}

The dimensional reduction of IIB supergravity on a circle produces 9d maximal supergravity. We can summarize the reduction of the relevant bosonic field content as follows, where we do not consider scalars, as they do not have zero-modes on AdS$_2$ and are hence irrelevant to the log-correction calculation. 

\begin{center}
\begin{tabular}{ |c|c|c|c|c| } 
 \hline
 10d & $g_{\mu\nu}$ & $B_{\mu\nu}$ & $C_{\mu\nu}$ & $C_{\alpha\beta\gamma\delta}^+$\\
 \hline\hline
9d & $g_{\mu\nu}$,$A_\mu$ & $B_{\mu}$,$B_{\mu\nu}$ & $C_{\mu\nu}, C_{
\mu}$ &$C_{\alpha\beta\gamma}$ or $C_{\alpha\beta\gamma\delta}$ \\ 
 \hline
\end{tabular}
\end{center}

The R-R self-dual four-form in IIB can be reduced to either a 4-form or a 3-form in 9d, and they can be Hodge dualized into each other. Although they are completely equivalent on-shell, they may produce a different one-loop result upon quantization \cite{DUFF1980179}. This issue has been discussed explicitly in \cite{Sen:2011ba}, and the question of which Hodge-duality frame one should use in the computation is related there to an ensemble choice.
The correct ensemble to match with the microscopic computations of the kind we do corresponds to using the 4-form in 9d, and we verify that by performing the 9d calculation before orbifolding and showing that the answer agrees with the elliptic genus in the $K3$ case \cite{Sen:2012cj}. See Appendix \ref{app:9dcomputation}.

The strategy that we adopt for computing logarithmic corrections has been developed in \cite{Sen:2012cj,Bhattacharyya:2012ye}. After dimensional reduction on $S^1$, we obtain \emph{locally} a product manifold $\left(AdS_2\times S^3\right)\times K3,$ where we use parentheses to distinguish dimensions where the characteristic scale becomes large in the scaling limit
\begin{equation}\label{eq:scalinglimit}
Q_1\sim Q_5\sim n\sim \Lambda,\quad r_h\sim (\sqrt{Q_1 Q_5 n})^{\frac{1}{3}}\sim\Lambda^{\frac{1}{2}},\qquad \Lambda \to \infty.
\end{equation}
Here $\Lambda$ is a dimensionless scaling constant, and $r_h$ is the horizon radius. Note that in this limit the characteristic scale of AdS$_2$ and $S^3$ grows as $\Lambda^{1/2}$ \cite{Sen:2012cj}, whereas the characteristic scale of $K3$ remains fixed. 

The logarithmic correction originates from the one-loop determinant of the 9d supergravity, and it can be captured by computing the heat kernel $K(\tau)$ of the associated kinetic operator, $\mathcal{A}$. In our case, the 9d manifold locally factorize into the large part, $AdS_2\times S^3$, and the small part, $K3.$ The metric under the scaling limit can be written as 
\begin{equation}
    ds^2=\Lambda g_{\mu\nu}^{(0)}dx^\mu dx^\nu +g_{ab}^{(1)}dy^a dy^b,
\end{equation}
where $(x^\mu,y^a)$ are respectively coordinates on $AdS_2\times S^3$ and $K3.$ Note that $g_{\mu\nu}^{(0)}, g_{ab}^{(1)}$ do not have $\Lambda$ dependence. The heat kernel can be schematically written as 
\begin{equation}
\begin{split} K(\tau)&=\sum_{m,n} e^{-(\kappa_m+\kappa_n')\tau}\ket{\kappa_m, \kappa_n'}\bra{\kappa_m, \kappa_n'}\\&=\sum_{m,\kappa_n'=0}e^{-\kappa_m\tau}\ket{\kappa_m, 0}\bra{\kappa_m, 0}+\sum_{m,\kappa_n'\neq 0}e^{-(\kappa_m+\kappa_n')\tau}\ket{\kappa_m, \kappa_n'}\bra{\kappa_m, \kappa_n'}=:K_0(\tau)+K_1(\tau),
\end{split}
\end{equation}
where $\kappa$ and $\kappa'$ are eigenvalues of $\mathcal{A}$ on AdS$_2\times S^3$ and $K3$ respectively, while $\ket{\kappa_m,\kappa'_n}$ are the subset of the eigen-vectors (or eigen-functions) of $\mathcal{A}$ on AdS$_2\times S^3\times K3$ that are invariant under the $\mathbb{Z}_2$ action. In the scaling limit (\ref{eq:scalinglimit}), $\kappa_m$ scales as $\frac{1}{\Lambda}$ whereas $\kappa_n'$ is of order $\Lambda^0.$ Thus to the leading order, $K_1(\tau)$ remains finite as $\Lambda \to \infty,$ whereas $K_0(\tau)$ diverges. The logarithmic correction we seek is therefore associated to $K_0(\tau).$

The integration over the non-zero modes of $\mathcal{A}$ contributes as \cite{Bhattacharyya:2012ye}
\begin{equation}\label{eq:heatkernel}
\begin{split}
    -\frac{1}{2}\log\text{det}'\mathcal{A}&=\frac{1}{2}\int_\epsilon^\infty \frac{d\tau}{\tau}\left(\text{Tr}K(\tau)-n^0_{\mathcal A}\right)\\&=\frac{1}{2}\int_\epsilon^\infty \frac{d\tau}{\tau}\left(\text{Tr}K_0(\tau)-n^0_{\mathcal A}\right)+\frac{1}{2}\int_\epsilon^\infty\frac{d\tau}{\tau}\text{Tr}K_1\left(\tau\right).
\end{split}
\end{equation}
The integral over $\tau$ is divergent at short distances, and we have inserted a UV cut-off $\epsilon.$ Also, following \cite{Bhattacharyya:2012ye} we used $\text{det}'$ to denote the determinant \emph{without} zero modes to make it well defined, and that corresponds to the subtraction of $n^0_{\mathcal A}$ from $\text{Tr}K\left(\tau\right).$

Because of the scaling of $\kappa_m$, $\kappa'_n$, the first integral on the second line of (\ref{eq:heatkernel}) is purely a function of $\bar{\tau}=\frac{\tau}{\Lambda}$ and contains a diverging term as $\Lambda\to \infty,$ whereas the second integral on the second line is finite in the $\Lambda\to \infty$ limit. In particular, the trace of $K_0$ can be evaluated by first noting that it receives contributions from modes that under the $\mathbb{Z}_2$ action are either even on both AdS$_2\times S^3$ and $K3$, or odd on both, so that their product survives the $\mathbb{Z}_2$ projection. 
We therefore have
\begin{equation}
\begin{split}
    \mathrm{Tr}K_0&=n^{0+}_{K3}\mathrm{Tr}^+_{\mathrm{AdS}_2\times S^3}K_0 +n^{0-}_{K3}\mathrm{Tr}^-_{\mathrm{AdS}_2\times S^3}K_0\\
    &=n^{0}_{K3/\mathbb{Z}_2}\mathrm{Tr}_{\mathrm{AdS}_2\times S^3/\mathbb{Z}_2}K_0+(n^0_{K3}-n^0_{K3/\mathbb{Z}_2})\big(\mathrm{Tr}_{\mathrm{AdS}_2\times S^3}K_0- \mathrm{Tr}_{\mathrm{AdS}_2\times S^3/\mathbb{Z}_2}K_0\big).\label{eq:eeoo}
\end{split}
\end{equation}
Here $n^0_{K3}$ and $n^0_{K3/\mathbb{Z}_2}$ denote the number of zero modes on $K3$ and $K3/\mathbb{Z}_2$. They are associated to the Betti numbers via Hodge theory. The superscripts $+/-$ denote whether the modes are even/odd under the $\mathbb{Z}_2$ action. In going to the second line we have used $n^{0+}_{K3}=n^{0}_{K3/\mathbb{Z}_2}$, as well as $n^{0}_{K3}=n^{0+}_{K3}+n^{0-}_{K3}$ and $\mathrm{Tr}^-=\mathrm{Tr}-\mathrm{Tr}^+$. Now we appeal to the expansion~\cite{Vassilevich:2003xt}
\begin{equation}\label{eq:heatkernel_expansion}
    \text{Tr}_{M_5}K_0(\tau)=\sum_{n=0}^\infty\frac{1}{\left(4\pi\right)^{\frac{5}{2}}}\tau^{\frac{n-5}{2}}\int \mathrm{d}^5x \sqrt{g} a_n(x,x),
\end{equation}
where $a_{n}(x,x)$ are known as the Seeley–De~Witt coefficients, and are local geometric invariants constructed from the 9d metric, the curvature tensor, the field strengths of gauge fields, and their covariant derivatives. In particular, $a_{n}(x,x)$ has length dimension $n,$ and thus $a_n(x,x)$ vanishes for odd $n$ because there is no local geometric invariant that can carry an odd number of derivatives.\footnote{We assume that the orbifold singularity, which occurs on an $S^1$ ($\times$AdS$_2$) where the $z=0$ plane intersects the $S^3$, does not invalidate our argument.
We provide evidence supporting this assumption in Appendix~\ref{app:heatkernel} by illustrating the heat-kernel computation for a scalar field on $S^3/\mathbb{Z}_2$, and leave a thorough investigation of this point for other types of fields to future work.} Writing in terms of $\bar{\tau}=\frac{\tau}{\Lambda}$ in order to identify the logarithmic term, we obtain 
\begin{equation}
\begin{split}
    -\frac{1}{2}\log\text{det}'A=\frac{1}{2}\int_{\epsilon/\Lambda}^\infty \frac{d\bar{\tau}}{\bar{\tau}}\left(\sum_{n=0}^\infty\frac{1}{\left(4\pi\right)^{\frac{5}{2}}}\bar{\tau}^{\frac{n-5}{2}} \Lambda^{\frac{5-n}{2}}\int \mathrm{d}^5x \sqrt{g} \tilde{a}_n(x,x)- n^0_{\mathcal A} \right).\label{eq:heatKerFormula}
\end{split}
\end{equation}
Here $\tilde{a}$ is the coefficient arising from the heat-kernel expansion of the combination on the second line of (\ref{eq:eeoo}), assuming that $\mathrm{AdS}_2\times S^3$ and $\mathrm{AdS}_2\times S^3/\mathbb{Z}_2$ have their own separate well-defined heat-kernel expansions.
The $\log \Lambda$ term can only arise from the $n=5$ term in the sum, which vanishes because $a_5$ should be zero for both AdS$_2\times S^3$ and AdS$_2\times S^3/\mathbb{Z}_2$. Thus in the path integral, the integration over non-zero modes contains the logarithmic correction:  
\begin{equation}\label{eq:logterm}
    -\frac{n^0_{\mathcal A}}{2}\int^\infty_{\frac{\epsilon}{\Lambda}} \frac{d\bar{\tau}}{\bar{\tau}}=-\frac{n^0_{\mathcal A}}{2}\log\frac{\Lambda}{\epsilon}+\mathcal{O}\left(1\right)\sim -\frac{n^0_{\mathcal A}}{2}\log \Lambda+ \dots
\end{equation}
We can also understand the presence of $\Lambda$ in (\ref{eq:logterm}) as being due to the presence of an infrared cut-off, because the logarithmic term is IR divergent too. Denoting the large distance cut-off by $r_0,$ the log correction would be of the form $\log\frac{r_0^2}{\epsilon}.$ Under the scaling limit (\ref{eq:scalinglimit}), $r_0$ must scale like the characteristic scale of AdS$_2$ and $S^3$, namely $r_0 \to r_0 \Lambda^{\frac{1}{2}},$ and thus produces the $\log \Lambda$ term in \eqref{eq:logterm}. We see indeed that $\log \Lambda$ is only sensitive to the IR effects, but is independent of the explicit IR cut-off $r_0$.

Taking into account the proper (fermionic vs bosonic) sign of the determinant, we thus find the integral over non-zero modes of $\mathcal{A}$ in the partition function contributes as 
\begin{equation}
    -\left(-1\right)^F n_{\mathcal A}^0\log \Lambda^{\frac{1}{2}}.
\end{equation}

One may want to consider alternatively a direct 10d calculation. In 10d the coefficient $a_{10}(x,x)$ is non-vanishing, and one might naively expect a $\log \Lambda^{\frac{1}{2}}$ term arising from $a_{10}$ in a 10d analog of \eqref{eq:heatKerFormula}. The above 9d discussion makes it clear however that only the large part of the geometry (i.e. the $AdS_2\times S^3$ part) plays a role in the relevant heat-kernel computation. Therefore a 10d calculation of the heat-kernel should proceed in parallel with the 9d computation above, yielding the same answer. Here we do not pursue the 10d calculation in detail, partly because of the difficulty in quantizing the 10d self-dual 4-form.


We now move on to the contributions from performing the path integral over the space of zero modes. In particular, they do not completely cancel with the ghost contributions, because of the existence of normalizable gauge transformations with non-normalizable gauge parameters in AdS. This phenomenon is a particular feature of even dimensional Anti-de Sitter space (or more generally of even-dimensional conformally compact manifolds \cite{ALBIN20071}). Let us assume integration over such zero modes produces a $\Lambda^{(-1)^F \frac{\beta\,  n^0_{\mathcal A}}{2}}$ factor in the path integral. Different from the cases in \cite{Sen:2012cj,Bhattacharyya:2012ye}, our scaling parameter $\Lambda$ is not an overall scale. Consider, for example, the properly normalized measure for a one-form: 
\begin{equation}
\int [D A_M] \exp\left(-\Lambda^{\frac{5}{2}}\int d^9x\sqrt{g^{(0)}}\left(\frac{1}{\Lambda} g^{(0) \mu \nu}A_{\mu}A_{\nu}+g^{(1) a b}A_{a}A_{b}\right)\right)=1,
\end{equation}
Then consider the path integral as being over values of $A_M(x)$ at each point, 
\begin{equation}
    \prod_x \prod_{\mu}\int dA_{\mu}(x) \exp\left(-\Lambda^{\frac{3}{2}} g^{(0) \mu \nu}(x)A_{\mu}(x)A_{\nu}(x) \Delta S_0\right)\prod_{a}\int dA_{a}(x) \exp\left(-\Lambda^{\frac{5}{2}} g^{(1) a b}(x)A_{a}(x)A_{b}(x) \Delta S_1\right),
\end{equation}
where $\Delta S_{0}$ and $\Delta S_{0}$ are the infinitesimal volume elements of $AdS_2\times S^3/\mathbb{Z}_2$ and $K3/\mathbb{Z}_2$ respectively. As in \cite{Sen:2012cj,Bhattacharyya:2012ye}, it is enough to consider the $AdS_2$ part as the non-normalizable gauge parameters only arise there. Thus when integrating over each $AdS_2$ zero mode, the properly normalized measure is $d(\Lambda^{\frac{3}{4}}A_\mu).$ Similarly, one can argue that the $\beta$ coefficient for each field is only sensitive to the scaling in $AdS_2.$

Thus we conclude that 
the computation of $\beta$ is only sensitive to the number of `large dimensions' under the scaling limit. In $D$ dimensions, the $\beta$ coefficients for various fields are (see e.g. \cite{Liu:2017vll})
\begin{equation}
    \beta_{\mathrm{graviton}}=D/2,\quad \beta_{\mathrm{gravitino}}=D-1,\quad \beta_{A_p}=D/2-p.\label{eq:betas}
\end{equation}
Based on the reasoning in the previous paragraph, we set $D=5$ even though we are performing the gravitational path integral in 9 dimensions. The $\beta$ factors are hence as in the following table.

\begin{center}
\begin{tabular}{ |c|c|c|c| } 
 \hline
gauge fields & metric & 4-form field & gravitino \\ 
\hline
 $\frac{3}{2}$ & $\frac{5}{2}$ & $-\frac{3}{2}$ & 4 \\ 
 \hline
\end{tabular}
\end{center}

Combining the contributions from non-zero modes and zero modes, the logarithmic contribution of a field with kinetic operator $\mathcal{A}$ can be written as 
\begin{equation}
    (-1)^F (\beta-1)n^0_{\mathcal A}\log \Lambda^{1/2}. \label{eq:macLogMaster}
\end{equation}

For a form field $A_p$, ghost contributions modify the above formula to
\begin{equation}
    \sum_{j=0}^p (-1)^j (\beta_{A_{p-j}}-j-1)n^0_{A_{p-j}}\log\Lambda^{1/2}.
\end{equation}
The graviton also has ghosts, but they do not contribute to the log correction; see e.g. \cite{Liu:2017vll}.

We do not consider spin-1/2 fermions because, just like scalars, they do not yield zero-modes on AdS$_2$. For the remaining fields the number of zero-modes on AdS$_2$ is \cite{Liu:2017vll}
\begin{equation}
    n^0_{\mathrm{graviton}}=-3,\quad n^0_{\mathrm{gravitino}}=-2,\quad n^0_{A_1}=-1.\label{eq:AdS2zeros}
\end{equation}

The coefficient of $\log\Lambda^{1/2}$ in the logarithmic correction  hence becomes
\be
\begin{split}
&\underbrace{-(\frac{3}{2}-1)\times 3}_{\text{gauge fields }A_1,B_1,C_1} -\underbrace{(\frac{5}{2}-1)\times (3+2)}_{\text{metric}}
+\underbrace{(\frac{3}{2}-2)\times 2}_{\text{1-ghosts of $C_2,B_2$}}+\underbrace{(4-1)\times 2}_{\text{gravitino}}-\underbrace{(-\frac{3}{2}-1)}_{\text{$\tilde{C}_4$}}\\&+\underbrace{(\frac{3}{2}-2)\times 10}_{\text{vectors from 3-ghost of $\tilde{C}_4$}}+\underbrace{(\frac{3}{2}-4)}_{\text{1-ghost of $\tilde{C}_4$}}=-9.\\
\end{split}\label{eq:logmatch}
\ee
Note that the contributions in the metric term come from the 2d metric and 2 gauge fields arising from the 2 $U(1)$ isometries of $(S^3\times K3)/\mathbb{Z}_2$, and the contribution in the three-form term comes from the 10 gauge fields arising from the 10 two-cycles in $(S^3\times K3)/\mathbb{Z}_2$. (See Appendix~\ref{app:hodge} for topological information on $(S^3\times K3)/\mathbb{Z}_2$.) Also, since we only have half the supersymmetry compared with Sen's $K3$ case, the gravitino contribution is half as much. 

The coefficient $-9$ in (\ref{eq:logmatch}) exactly matches with the $-9$ in  (\ref{eq:logCor}). 

\subsubsection*{5d calculation}



We can also consider performing the gravitational path integral on the five large dimensions of spacetime. These are the dimensions with characteristic scale $\Lambda^{1/2}$ as $\Lambda\to\infty$. As discussed above, the 5d picture arises not from a conventional KK reduction (since the $\mathbb{Z}_2$ mixes the large $S^3$ with the small $K3$), but from a low-energy ($E\ll 1/R_{S^1},1/\sqrt{\alpha'}$) perspective. The 5d near-horizon geometry is AdS$_2\times (S^3/\mathbb{Z}_2)$, which is singular because the $\mathbb{Z}_2$ leaves an $S^1$ subset of the $S^3$ fixed. This is an orbifold singularity, and string theory computation of quantum effects on backgrounds with such orbifold singularities is well-understood. See in particular \cite{Ardehali:2013xya}, where it is shown that to obtain results in agreement with microscopic calculations, quantum corrections due to massless string states arising from the twisted sector localized near the singularity are essential. In the present setting there are no massless twisted string states, because the orbifold singularity is resolved at the scale $1/\sqrt{\alpha'}$ where the $K3$ and the free action of the $\mathbb{Z}_2$ on it become visible. Said differently, the twisted strings have masses of order $1/\sqrt{\alpha'}$ (see the discussion section below) and hence do not contribute to the log correction that we are after. Therefore we can adopt a 5d perspective and still work only with the massless modes arising from 9d supergravity as discussed above.

We consider IIB supergravity reduced on $K3$ (as the $K3$ zero-modes are trivially invariant under the $\mathbb{Z}_2$) and then reduce on $S^1$. We shall ignore scalars and fermions, and only keep differential forms, gravitinos, and the graviton, as they can potentially have zero modes. The relevant 5d field content is summarized in the following table.
\begin{center}
\begin{tabular}{ |c|c|c|c|c| } 
 \hline
 10d & $g_{\mu\nu}$ & $B_{\mu\nu}$ & $C_{\mu\nu}$ & $C_{\alpha\beta\gamma\delta}^+$\\
 \hline\hline
5d & $g_{\mu\nu}$,$A_\mu$ & $B_{\mu}$,$B_{\mu\nu}$ & $C_{\mu}, C_{\mu\nu}$ &$10\times C_\alpha$ or $10\times C_{\alpha\beta}$ \\ 
 \hline
\end{tabular}
\end{center}
In the above table, the 1-forms from the 10d graviton and 2-forms come from wrapping over the $S^1$, and we obtain 10 additional 1-forms (or dual 2-forms) from the self-dual 4-form in 10d as it can wrap on the 10 distinct 2-cycles of the Enriques surface (see Appendix~\ref{app:hodge}). 


Note that in 5d, vectors are dual to 2-forms, and from the 5d perspective it is natural to dualize the matter in terms of vectors (which are the ``lower'' forms) in 5d supergravity. This is exactly what is done in the standard $K3$ case in \cite{Sen:2012cj}, and we do the same in the orbifolded geometry. Thus we conclude that, in 5d we have one graviton and $15$ vector fields. Using the formulas \eqref{eq:macLogMaster}--\eqref{eq:AdS2zeros}, we obtain the coefficient of $\log\Lambda^{1/2}$ in the logarithmic correction to be
\be
\begin{split}
&\underbrace{-(\frac{3}{2}-1)\times 15}_{\text{gauge fields}} -\underbrace{(\frac{5}{2}-1)\times (3+2)}_{\text{metric}}
+\underbrace{(4-1)\times 2}_{\text{gravitino}}=-9,\\
\end{split}\label{eq:logcorrection5d}
\ee
again in exact agreement with (\ref{eq:logCor}). This result could in fact be inferred from the calculation in 5d of $\mathcal{N}=2$ supergravity with $n_v$ vector fields (including the graviphoton) done in \cite{Sen:2012cj}, by setting $n_v=15$ (as the above table indicates) in that work.

We note that a different choice of Hodge duality frame for the form fields would result in a different answer from (\ref{eq:logcorrection5d}). However, the Hodge duality frame corresponds to the boundary conditions of the gravitational path integral, which is related to an ensemble choice. One has to choose the right ensemble to match with the one used in a given microscopic computation, as commented in \cite{Sen:2011ba}. Here we have not investigated the ensemble question directly, but rather fixed the Hodge-duality frame just as in the well-studied $K3$ setting---from which our ES geometry descends via a $\mathbb{Z}_2$ orbifold.

We emphasize that what we referred to as ``5d calculation'' or ``9d calculation'' above, both involve essentially just zero-mode counting on AdS$_2$, as well as fixing various scaling factors $\beta_r$ which are the same in the two approaches. The two calculations differ only because $i)$ the treatment of the non-zero modes is slightly different (with that in the 9d calculation seemingly more thorough), and $ii)$ the choice of the Hodge duality frame can be made either in 9d (where we fixed the frame by noting that it is the 4-form that gives the correct answer for the standard $K3$ case, see Appendix~\ref{app:9dcomputation}) or in 5d (where we fixed the frame as in \cite{Sen:2012cj} by dualizing the 2-forms into vectors).

\subsection*{Adding angular momentum}

In the ES case, the macroscopic log corrections are the same for $J=0$ and $J\neq 0$. This is because unlike in the $K3$ case \cite{Sen:2012cj}, adding angular momentum does not break the isometry of our near-horizon geometry any further than the $\mathbb{Z}_2$ orbifold does. As a result, on the macroscopic side the near-horizon zero-mode content does not change for nonzero $J$, and hence neither does the log correction. 

The fact that the near-horizon symmetry does not get reduced by addition of angular momentum also implies that on the microscopic side the same ensemble with degeneracy $\tilde{d}$ remains appropriate. Although our microscopic computation in Section~\ref{subsec:microCounting} assumed $J=0$, it is straightforward to check that adding angular momentum does not change the microscopic result $-9\ln\Lambda^{1/2}$, and hence the match with the macroscopic side remains intact. To see this explicitly, note that reinstating $J$ in (\ref{eq:deg&taus}) we get
\begin{equation}
\tilde{d}_{\mathrm{micro}}^{\mathrm{ES}}(n,Q_1,Q_5,J)\simeq\int \mathrm{d}\tau_1\int
\mathrm{d}\tau_2\ e^{\frac{\pi
}{\tau_2}(Q_1Q_5(\tau_1^2+\tau_2^2)+n-J\tau_1)}\eta(-\tau_1+i\tau_2)^{-12}\eta(\tau_1+i\tau_2)^{-12}\tau_2^{-8}.\label{eq:deg&tausWithJ}
\end{equation}
The shift in the exponent of the exponential, compared with (\ref{eq:deg&taus}), changes the critical values of $\tau_{1,2}$ to $\tau_1=J/2Q_1Q_5$ and $\tau_2=\sqrt{(n-J^2/4Q_1Q_5)/Q_1Q_5}$. However, $\tau_2$ remains of order $\Lambda^{-1/2}$, and the effective widths of the two integrals remain of order $\Lambda^{-5/4}$. We thus arrive at
\begin{equation}
\tilde{d}_{\mathrm{micro}}^{\mathrm{ES}}(n,Q_1,Q_5,J)\simeq
e^{2\pi\sqrt{Q_1Q_5n-J^2/4}}\Lambda^{-9/2},\label{eq:degFinalWithJ}
\end{equation}
with the same logarithmic correction that we had for $J=0$, as claimed.


\section{Summary and discussion}\label{sec:open}

In this paper we proposed D1-D5 systems realizing the $\mathcal{N}=(2,2)$ AdS$_3$/CFT$_2$ dualities of \cite{Eberhardt:2017uup} in string theory.

The two-charge systems (with $Q_1,Q_5$) have near-horizon geometry AdS$_3\times (S^3\times M_4)/G$, dual to the vacuum state of the boundary CFT. We showed in Section~\ref{sec:branes} that the Brown-Henneaux central charge of the near-horizon AdS$_3$ and the quantum correction to it match respectively the leading and the subleading central charge of the boundary sigma model.

The three-charge systems (with $Q_1,Q_5,n$) yield black branes with near-horizon geometry AdS$_2\times S^1 \times (S^3\times M_4)/G$, dual to an ensemble of excited states in the boundary CFT. We argued in Section~\ref{sec:indices} that the Bekenstein-Hawking entropy of the black branes matches the Cardy entropy of the ensemble of excited states in the boundary sigma model. In the ES case---where $M_4=K3$ and $G=\mathbb{Z}_2$---we also derived the Bekenstein-Hawking entropy from the CFT elliptic genus. The elliptic genus moreover yields a logarithmic correction to the Bekenstein-Hawking entropy, which we reproduced macroscopically through one-loop computations on the near-horizon background. In the HS cases---where $M_4=T^4$---the elliptic genus vanishes,
so the analogous microscopic computations require studying modified supersymmetric indices (as in \cite{Maldacena:1999bp}), which we have left to future work.

\subsection*{The constraint on $Q_5$}

A novel aspect of the $\mathcal{N}=(2,2)$ brane systems, unprecedented in the standard $(4,4)$ context, is the constraint on the number of D5 branes. Let us begin by discussing the constraint in the ES case where the picture is most coherent.

The Coulomb branch considerations of Subsection~\ref{subsec:orbFiber} imply that in the ES case $Q_5$ should be odd, otherwise a non-standard Coulomb-branch decoupling is needed to get rid of the extra moduli. The sigma model considerations of Subsection~\ref{subsec:sigma} further reinforced the odd-$Q_5$ constraint, because the target space \eqref{eq:ESsigma} following from the standard arguments does not make sense for even $Q_5$. While in principle it is possible that some non-standard modifications can lead to a consistent duality in the even-$Q_5$ case, here we assume $Q_5$ is odd to stay within the standard AdS$_3$/CFT$_2$ framework.

With odd $Q_5$ the inflow formula (\ref{eq:Q1}) implies that $Q_1$ should be in $\mathbb{Z}+1/2$. In other words, in the ES case we have a shifted Dirac quantization between the D1 and D5 charges, as the latter is in $\mathbb{Z}$. Such shifted quantizations are usually associated to anomalies; see e.g. \cite{Witten:1996md,Hsieh:2020jpj,Moore:1999gb,Witten:2000cn}. In fact the somewhat unusual $Q_1\in\frac{1}{2}\times\mathbb{Z}$ follows readily from the I-brane inflow of \emph{local} anomalies \cite{Green:1996dd} referred to in Section~\ref{sec:branes}. However, an anomaly explanation of the stronger condition that  $Q_1\in\mathbb{Z}+1/2$ seems to require considerations of \emph{global} anomalies.

A relatively well-understood source of such global anomalies is world-volume fermions. Global fermion anomalies can shift Dirac quantization conditions via eta-invariants; see e.g. \cite{Tachikawa:2018njr}. For the D-brane configuration underlying the ES duality the relevant eta-invariant\footnote{Compare with Section~2.4 of \cite{Tachikawa:2018njr}, with $p=1$ and $q=5$.} is that of the space supporting the RR flux sourced by the D1 branes, namely $\eta((S^3\times K3)/\mathbb{Z}_2)$. It quantifies the global anomaly of the fermions in the world-volume of the D5s. If we had
\begin{equation}
    \eta((S^3\times K3)/\mathbb{Z}_2)\overset{?}{\in}\mathbb{Z}+1/2,\label{eq:eta?}
\end{equation}
the shifted Dirac quantization in the ES case would be perfectly explained by such global fermion anomalies \cite{Tachikawa:2018njr}. However, the following argument implies that (\ref{eq:eta?}) is \emph{not true}.\footnote{We are indebted to K.~Yonekura for instructive correspondences on this point.} We can compute $\eta((S^3\times K3)/\mathbb{Z}_2)$ via the equivariant APS index theorem in Appendix~D.2 of \cite{Hsieh:2020jpj}. Let $Y=S^3\times K3$. We have
\begin{equation}
    \eta(Y/\mathbb{Z}_2)=\frac{1}{2}\big(\eta(Y,1)+\eta(Y,G)\big),
\end{equation}
with $G$ the generator of the $\mathbb{Z}_2.$ However, we have $\eta(Y,1)=\eta(Y)$, which is zero (because $\eta(M\times N)=\eta(M)\tau(N)+\tau(M)\eta(N)$, and $\eta(S^3)=\eta(K3)=0$). So we only need to compute $\eta(Y,G)$. Now let $Z=D^4\times K3$, so that $Y=\partial Z$. Since $G$ does not have a fixed point on $Z$, the equivariant index theorem (Eq.~(D.22) of \cite{Hsieh:2020jpj}) implies $\eta(Y,G)=\mathrm{index}(Z,G)$. The latter is defined as a trace over APS zero-modes (Eq.~(D.20) of \cite{Hsieh:2020jpj}) and should vanish because $D^4$ does not have APS zero modes. It follows that $\eta((S^3\times K3)/\mathbb{Z}_2)=0$.\footnote{Incidentally, a similar argument applies to the simplest HS case as well, implying $\eta((S^3\times T^4)/\mathbb{Z}_2)=0$.}

Since (\ref{eq:eta?}) is not true, the D5-brane global fermion anomalies can not explain the shifted Dirac quantization. Another possibility is that the required anomaly arises from the bosonic fields in the D5-brane world-volume; while for D3, D4, and M5 branes some such anomalies have been studied in \cite{Hsieh:2020jpj}, the case of D5 branes appears to be more difficult. It may also be that properly addressing the shifted quantization of $Q_1$ in the ES case requires taking the $B$-field vev into account, which we have neglected throughout most of this work. Also, since ES is not spin, there is a Freed-Witten \cite{Freed:1999vc} shifted quantization in D5-brane worldvolume gauge fluxes, which may have an interplay with the D1 and D5 brane charges along the lines studied in \cite{Aharony:2009fc}.

Although we have not found an explanation for the odd-$Q_5$ constraint in the ES case from anomalies in the D-brane picture, an explanation in the S-dual picture seems to follow from an argument similar to the one provided in \cite{Datta:2017ert} for a $\mathcal{N}=(2,2)$ duality studied in that work. There a D-brane picture is lacking at present, but in the available picture (presumably S-dual to a D-brane setting) the NSNS flux through the three-cycle of the bulk geometry AdS$_3\times (S^3\times T^2)/\mathbb{Z}_2 \times T^2$ should be odd, otherwise the twisted-sector strings would break the spacetime $\mathcal{N}=(2,2)$ supersymmetry (see Section~6.2 of \cite{Datta:2017ert} for the precise argument).
In our ES case, after S-duality, odd $Q_5$ translates to a similar odd NSNS flux constraint, which seems to be analogously required again for the ``twisted sector'' to preserve $(2,2)$ spacetime supersymmetry.
The main difference is that in the ES duality of our interest, the $\mathbb{Z}_2$ orbifold acts freely on $S^3\times K3$. So the ``twisted'' strings, which arise from open strings on the covering space that are stretched between those points of the $K3$ identified by the $\mathbb{Z}_2$ quotient, are actually closed strings wrapping the torsion cycle of the orbifolded geometry (note that $H_1\big((S^3\times K3)/\mathbb{Z}_2;\mathbb{Z}\big) =\mathbb{Z}_2$). Assuming, as usual, that the size of the $K3$ is of order $\sqrt{\alpha'}$ (see Eq.~(\ref{eq:volK3}) below), such closed strings wrapping the non-contractible loop would have masses of order $\mathrm{length}\times\mathrm{tension}=\sqrt{\alpha'}\times1/\alpha'=1/\sqrt{\alpha'}$. They would hence be absent from the low-energy theory in the $\alpha'\to0$ limit. Nevertheless, away from the $\alpha'\to0$ limit the stretched strings are present, and hence in the ES case the odd-$Q_5$ constraint can be thought of as a necessary condition for supersymmetry at non-zero $\alpha'$. 
In the light of the AdS$_3$/CFT$_2$ relation \cite{Maldacena:1997re}

\begin{equation}
\mathrm{vol}(K3)\sim \alpha'^2 Q_1/Q_5,\label{eq:volK3}   
\end{equation}
it is indeed reasonable to have the non-zero-$\alpha'$ effects on the string-sized ES be correlated with the constraints arising at finite $Q_5$---here the constraint being that $Q_5$ is odd.

A similar ``twisted-sector SUSY'' argument seems to apply to the (S-dual of the) simplest HS case---where $G=\mathbb{Z}_2$---as well, implying that $Q_5$ should be odd in that case too. (Analogous arguments can presumably be given in the other HS cases as well, but we do not attempt that here.) This fits nicely with our Coulomb branch considerations in Subsection~\ref{subsec:orbFiber}. However, unlike in the ES case, we did not see any $Q_5$ constraints arise from the associated sigma models (\ref{eq:HSsigma}) in the HS cases. So the picture is not as clear in these cases.

In conclusion, constraints on fluxes appear to be a common feature in the new $\mathcal{N}=(2,2)$ AdS$_3$/CFT$_2$ dualities \cite{Datta:2017ert,Eberhardt:2017uup}, but one that is yet to be properly understood.


\begin{acknowledgments}
We would like to thank L.~Eberhardt for collaboration during early stages of this project, providing us with the content of Appendix~\ref{app:hodge}, and several helpful correspondences during final stages of this work. We are also thankful to O.~Aharony, N.~Benjamin, O.~Bergman, A.~Charles, A.~Gustavsson, C.~Keller, I.~Klebanov, D.~Kutasov, F.~Larsen, J.~Minahan, U.~Naseer, L.~Pando~Zayas, A.~Pittelli, S.~Sheikh-Jabbari, Y.~Tachikawa, K.~Yonekura, and G.~Zafrir for helpful discussions and correspondences on related topics. This work was supported in part by ERC under the STG grant 639220, by Vetenskapsr\r{a}det under the grant 2018-05572, and by the ERC consolidator grant 681908, “Quantum black holes: A macroscopic window into the microstructure of gravity”.

\end{acknowledgments}

\appendix

\section{Elliptic genus for supergravity on $\mathrm{AdS}_3\times (S^3\times
K3)/\mathbb{Z}_2$}\label{app:1}

To compute the supergravity elliptic genus we need the KK supergravity spectrum on AdS$_3\times (S^3\times
K3)/\mathbb{Z}_2$. To obtain the latter, we start from the spectrum on AdS$_3\times S^3\times
K3$, twisted by a formal variable $\alpha$, satisfying $\alpha^2=1$, that keeps track of the $\mathbb{Z}_2$ parity of the states (see Section~4.1 of \cite{Eberhardt:2017uup}):
\begin{equation}
\begin{split}
&(0,0)+(11+10\alpha)(\frac{1}{2},\frac{1}{2})^\alpha_S+\alpha(0,1)^\alpha_S+\alpha(1,0)^\alpha_S+\alpha(\frac{1}{2},\frac{3}{2})^\alpha_S+\alpha(\frac{3}{2},\frac{1}{2})^\alpha_S\\
&\bigoplus_{m\ge3}
(12+10\alpha)(\frac{m-1}{2},\frac{m-1}{2})^\alpha_S+\alpha(\frac{m-1}{2},\frac{m+1}{2})^\alpha_S+\alpha(\frac{m+1}{2},\frac{m-1}{2})^\alpha_S\ .\label{eq:spectES}
\end{split}
\end{equation}
Here $(h,\bar{h})^\alpha_S$ denote the modified short
$\mathcal{N}=(4,4)$ representations \cite{Datta:2017ert,Eberhardt:2017uup}, which are the standard short $\mathcal{N}=(4,4)$ representations twisted by $\alpha$. Note that our notation
differs from that of \cite{Datta:2017ert,Eberhardt:2017uup} in that
instead of writing the SU($2$) dimensions inside the parentheses, we
have written the SL($2,\mathbb{R}$) quantum numbers of the lowest components
of the corresponding $\mathcal{N}=4$ multiplets.

To spell out the notation in (\ref{eq:spectES}) more explicitly, let us denote a state on the left sector by $|h,j\rangle$, where $h$ and $j$ are respectively the SL($2,\mathbb{R}$) and U(1)$_R$ quantum numbers. Then the standard short $\mathcal{N}=4$ representation $(h)_S$ on the left sector contains (see Appendix~A.1 of \cite{ArabiArdehali:2018mil})
\begin{equation*}
    \begin{tabular}{ c c c }
        $|h,h\rangle\qquad$ & $2\times |h+\frac{1}{2},h-\frac{1}{2}\rangle\qquad$ & $|h+1,h-1\rangle$\\ 
        $|h,h-1\rangle\qquad$ & $2\times |h+\frac{1}{2},h-\frac{3}{2}\rangle\qquad$ &  $|h+1,h-2\rangle$\\  
        $\vdots\qquad$ & $\vdots\qquad$ & $\vdots$\\    
        $|h,-h+1\rangle\qquad$ & $2\times |h+\frac{1}{2},-h+\frac{3}{2}\rangle\qquad$ & $|h+1,-h+2\rangle$\\
        $|h,-h\rangle\qquad$ & $2\times |h+\frac{1}{2},-h+\frac{1}{2}\rangle\qquad$ & $|h+1,-h+1\rangle.$
\end{tabular}
\end{equation*}
Its modified version $(h)^{\alpha}_S$ is then defined by the content (see Eq.~(C.3) of \cite{Eberhardt:2017uup})
\begin{equation*}
    \begin{tabular}{ c c c }
        $|h,h\rangle\qquad$ & $(1+\alpha)\times |h+\frac{1}{2},h-\frac{1}{2}\rangle\qquad$ & $\alpha|h+1,h-1\rangle$\\
        $\alpha|h,h-1\rangle\qquad$ & $(1+\alpha)\times |h+\frac{1}{2},h-\frac{3}{2}\rangle\qquad$ &  $|h+1,h-2\rangle$\\  
        $\vdots\qquad$ & $\vdots\qquad$ & $\vdots$\\    
        $\alpha^{2h-1}|h,-h+1\rangle\qquad$ & $(1+\alpha)\times |h+\frac{1}{2},-h+\frac{3}{2}\rangle\qquad$ & $\alpha^{2h-2}|h+1,-h+2\rangle$\\
        $\alpha^{2h}|h,-h\rangle\qquad$ & $(1+\alpha)\times |h+\frac{1}{2},-h+\frac{1}{2}\rangle\qquad$ & $\alpha^{2h-1}|h+1,-h+1\rangle.$
\end{tabular}
\end{equation*}
The idea is that once the content of the modified $\mathcal{N}=4$ multiplets in (\ref{eq:spectES}) are expanded as such, and all even powers of $\alpha$ are replaced with 1, we then set $\alpha=0$ to obtain the $\mathbb{Z}_2$-singlet spectrum of AdS$_3\times S^3 \times K3$, which is the desired spectrum on AdS$_3\times (S^3\times
K3)/\mathbb{Z}_2$.

For example, the vacuum $(0,0)$ in (\ref{eq:spectES}) is certainly $\mathbb{Z}_2$-singlet, as it is not multiplied by an odd power of $\alpha$, and hence survives the projection. Moving on to the second term in (\ref{eq:spectES}), it contains a tensor product of two copies of $(\frac{1}{2})^{\alpha}_S$, each containing $|\frac{1}{2},\frac{1}{2}\rangle,(1+\alpha)|1,0\rangle,\alpha|\frac{1}{2},-\frac{1}{2}\rangle$. Since it is multiplied by $(11+10\alpha)$, it yields 11 states of the form $|\frac{1}{2},\frac{1}{2}\rangle\otimes |\frac{1}{2},\frac{1}{2}\rangle$, as well as 10 states of the form $|\frac{1}{2},\frac{1}{2}\rangle\otimes |\frac{1}{2},-\frac{1}{2}\rangle$, another 10 states of the form $|\frac{1}{2},-\frac{1}{2}\rangle\otimes |\frac{1}{2},\frac{1}{2}\rangle$, and so on.

The computation of the supergravity elliptic genus $\mathcal{I}_R(q)$ becomes straightforward if we write out the spectrum (\ref{eq:spectES}) in
$\mathcal{N}=(2,2)$ language. In fact since on the right sector only
the chiral multiplets of $\mathcal{N}=2$ contribute to
$\mathcal{I}_R(q)$, we only need the anything$\times$chiral spectrum
of the theory. On the left sector we denote the long multiplets of
$\mathcal{N}=2$ (containing four states) by long$^{j}_{h}$, with $j$
the charge of the lowest-component state under $J_0$. On either
side, for labeling the short multiplets of $\mathcal{N}=2$ (containing two
states) we use only the SL($2,\mathbb{R}$) quantum number of the lowest-component state. See Appendix~A.1 of \cite{ArabiArdehali:2018mil} for background material on the representation theory involved.

The anything$\times$chiral spectrum in (\ref{eq:spectES}) reads
\begin{equation}
\begin{split}
&(0,0)+11(\mathrm{chiral}_{1/2},\mathrm{chiral}_{1/2})^{\mathcal{N}=2}+10(\mathrm{antichiral}_{1/2},\mathrm{chiral}_{1/2})^{\mathcal{N}=2}+(\mathrm{long}^0_{1},0)^{\mathcal{N}=2}\\
&+ \cdots. \label{eq:specN2ES}
\end{split}
\end{equation}

For the purpose of illustration, we compute the contribution of the
$(\mathrm{chiral}_{1/2},\mathrm{chiral}_{1/2})^{\mathcal{N}=2}$ and
$(\mathrm{long}^0_{1},0)^{\mathcal{N}=2}$ multiplets to
$\mathcal{I}_R(q)$.

The $\mathrm{chiral}_{1/2}$ multiplet on the right sector of
$(\mathrm{chiral}_{1/2},\mathrm{chiral}_{1/2})^{\mathcal{N}=2}$
contributes
\begin{equation*}
(-1)^{-2\bar{h}}=-1
\end{equation*}
to $\mathcal{I}_R(q)$. The $\mathrm{chiral}_{1/2}$ multiplet on the
left sector contributes
\begin{equation*}
(-1)^{2h}\frac{\sqrt{q}-q}{1-q}=-\frac{\sqrt{q}-q}{1-q},
\end{equation*}
where the $\sqrt{q}$ and $q$ in the numerator come from the lowest
weight state (with $h=1/2$ and $j=1/2$) and its supersymmetric
partner (with $h=1$ and $j=0$), while the $1-q$ in the denominator
comes from taking into account their $L_{-1}$ descendants. For the
combined multiplet we thus obtain
\begin{equation}
\mathcal{I}_R[(\mathrm{chiral}_{1/2},\mathrm{chiral}_{1/2})^{\mathcal{N}=2}]=\frac{\sqrt{q}-q}{1-q}.
\end{equation}

Next, we note that the vacuum on the right sector of
$(\mathrm{long}^0_{1},0)^{\mathcal{N}=2}$ contributes
\begin{equation*}
(-1)^{-2\bar{h}}=1
\end{equation*}
to $\mathcal{I}_R(q)$. The $\mathrm{long}^0_{1}$ multiplet on the
left sector contributes
\begin{equation*}
(-1)^{2h}\frac{q-2q^{3/2}+q^2}{1-q}=\frac{q-2q^{3/2}+q^2}{1-q},
\end{equation*}
where the $q$, $q^{3/2}$, and $q^2$ in the numerator come from the
lowest weight state (with $h=1$ and $j=0$), its supersymmetric
partners (with $h=3/2$ and $j=\pm1/2$), and the highest weight state
(with $h=2$ and $j=0$), while the $1-q$ in the denominator comes
from taking into account their $L_{-1}$ descendants. For the
combined multiplet we hence get
\begin{equation}
\mathcal{I}_R[(\mathrm{long}^0_{1},0)^{\mathcal{N}=2}]=\frac{q-2q^{3/2}+q^2}{1-q}.
\end{equation}
Summing up all such contributions in the spectrum (\ref{eq:specN2ES}) the result (\ref{eq:ESeg}) follows.

A more efficient derivation is possible using the so-called ``modified $\mathcal{N}=4$ characters''. For a representation $(h)_S^\alpha$ the modified character is (see Eq.~(C.3) of \cite{Eberhardt:2017uup})
\begin{equation}
    \chi_h^{\mathcal{N}=4,\alpha}(q,y)=\frac{ (\alpha q)^h}{1-q}\left(\chi_h(y/\alpha)-(\frac{q}{\alpha})^{\frac{1}{2}}(1+\alpha)\chi_{h-\frac{1}{2}}(y/\alpha)+q\chi_{h-1}(y/\alpha)\right),
\end{equation}
where
\begin{equation}
    \chi_\ell(y)=\frac{y^{-\ell-\frac{1}{2}}-y^{\ell+\frac{1}{2}}}{y^{-\frac{1}{2}}-y^{\frac{1}{2}}},
\end{equation}
is the SU(2) character. The contribution of a representation $(h,\bar{h})_S^{\alpha}$ to $\mathcal{I}_R(q,y)$ is then
\begin{equation}
    (-1)^{2(h-\bar{h})} \chi_h^{\mathcal{N}=4,\alpha}(q,y).
\end{equation}
Summing over the spectrum (\ref{eq:spectES}) yields an elliptic genus $\mathcal{I}^\alpha_R(q,y)$, from which the $K3$ supergravity elliptic genus is obtained by setting $\alpha=1$. That of ES is obtained as
\begin{equation}
    \frac{\mathcal{I}^{\alpha=1}_R(q,y)+\mathcal{I}^{\alpha=-1}_R(q,y)}{2},
\end{equation}
and in the limit $y\to1$ we get (\ref{eq:ESeg}).

\section{9d computation of the log correction for $\mathrm{AdS}_2\times S^3\times K3$}\label{app:9dcomputation}
Consider a 9d path integral on AdS$_2\times S^3\times K3$, with the IIB reduced on $S^1:$
\begin{center}
\begin{tabular}{ |c|c|c|c|c| } 
 \hline
 10d & $g_{\mu\nu}$ & $B_{\mu\nu}$ & $C_{\mu\nu}$ & $C_{\alpha\beta\gamma\delta}^+$\\
 \hline\hline
9d & $g_{\mu\nu}$,$A_\mu$ & $B_{\mu}$,$B_{\mu\nu}$ & $C_{\mu\nu}, C_{
\mu}$ &$C_{\alpha\beta\gamma}$ or $C_{\alpha\beta\gamma\delta}$ \\ 
 \hline
\end{tabular}
\end{center}
Keeping the four-form in 9d, the formalism outlined in Subsection~\ref{subsec:macroLog} gives the coefficient of $\log\Lambda^{1/2}$ in the logarithmic correction as 
\be
\begin{split}
&\underbrace{-(\frac{3}{2}-1)\times 3}_{\text{gauge fields}} -\underbrace{(\frac{5}{2}-1)\times (3+6)}_{\text{metric}}
+\underbrace{(\frac{3}{2}-2)\times 2}_{\text{1-ghosts of $C_2,B_2$}}+\underbrace{(4-1)\times 4}_{\text{gravitino}}-\underbrace{(-\frac{3}{2}-1)}_{\text{$\tilde{C}_4$}}\\&+\underbrace{(\frac{3}{2}-2)\times 22}_{\text{vectors from 3-ghost of $\tilde{C}_4$}}+\underbrace{(\frac{3}{2}-4)}_{\text{1-ghost of $\tilde{C}_4$}}=-15,\\
\end{split}\label{eq:logmatchK3}
\ee
which matches the result of \cite{Sen:2012cj}. The factor of $3$ in the gauge fields contribution is because there are three gauge fields $A_\mu,B_\mu,C_\mu$, while the $+6$ in the graviton contribution is because $S^3$ has a six-dimensional isometry group SU(2)$\times $SU(2) so the graviton yields six vector fields on AdS$_2$, and so on.

If we instead choose to keep the three-form in 9d, the result changes to $-27/2$, which does not match the microscopic calculation of \cite{Sen:2012cj}. This indicates that the correct ensemble to use to compare with the microscopic calculation corresponds to using the 4-form instead of the 3-form. We hence did the same in the $\mathbb{Z}_2$ orbifolded case in the main text.

\section{The heat kernel on $S^3/\mathbb{Z}_2$}\label{app:heatkernel}

The heat kernel calculation in the main text relied on the assumption that the singularity arising from the $\mathbb{Z}_2$ quotient does not introduce additional logarithmic corrections. In this appendix we consider the heat kernel of a scalar field on $S^3/\mathbb{Z}_2$, with the $\mathbb{Z}_2$ action having a fixed circle, and demonstrate the validity of the said assumption. This justifies the heat kernel computation in the main text for scalars. The discussion is similar to the $S^2/\mathbb{Z}_2$ case in \cite{Gupta:2013sva}.

It is convenient to use the coordinates 
\begin{equation}
    \mathrm{d}s^2=\mathrm{d}\psi^2+\cos^2\psi\, \mathrm{d}\theta_1^2+\sin^2 \psi\,  \mathrm{d}\theta_2^2 ,
\end{equation}
where $\psi\in[0,\frac{\pi}{2}], \theta_i\in [0,2\pi],$ and for simplicity we have taken the radius of $S^3$ to be 1. In these coordinates, a scalar field can be decomposed into $S^3$ harmonics that satisfy
\begin{equation}
    \Delta_{S^3}Y_{kmn}=-k(k+2)Y_{kmn},
\end{equation}
where $k\in \mathbb{Z}^+.$ Note that $Y_{kmn}\sim A_k(\psi) e^{i m \theta_1 + i n \theta_2},$ and $A_{kmn}(\psi)$ can be found explicitly: 
\begin{equation}
    \begin{split}
A_{kmn}(\psi)&= c_1 x^m(1-x^2)^{\frac{n}{2}}\,{}_2F_1\big(\frac{1}{2}(-k+m+n),\frac{1}{2}(2+k+m+n),1+m,x^2\big)\\
&+ c_2(-1)^mx^{-m}(1-x^2)^{\frac{n}{2}}\,{}_2F_1\big(\frac{1}{2}(-k-m+n),\frac{1}{2}(2+k-m+n),1-m,x^2\big),
    \end{split}
\end{equation}
with $x:=\cos \psi$. The regularity condition at $x=0$ requires we set $c_2=0$ for $m\geq 0$ and $c_1=0$ for $m<0.$ The regularity condition at $x=\pm 1$ amounts to 
\begin{equation}
    0\leq |m-n| \leq k,\,\  0\leq |m+n| \leq k,\,\  k-m-n\in 2\mathbb{Z}.
\end{equation}
For each $k$, there is a $(k+1)^2$ fold degeneracy, and thus the heat kernel is given by 
\begin{equation}
    \text{Tr}K_{S^3}\left(s\right)=\sum_{k=0}^{\infty} (k+1)^2e^{-s k(k+2)}.
\end{equation}

The $\mathbb{Z}_2$ action described in Section~\ref{subsec:pBrane} maps $\theta_2\to \theta_2+\pi.$ Such an action only leaves modes with even $n.$ For odd $k$ such an action reduces the degeneracy to $\frac{1}{2}(k+1)^2$, and for even $k$ to $\frac{1}{2}((k+1)^2+1).$ Thus for $S^3/\mathbb{Z}_2$ we find 
\begin{equation}
    \text{Tr}K_{S^3/\mathbb{Z}_2}\left(s\right)=\frac{1}{2}\text{Tr}K_{S^3}(s)+ \frac{1}{2}\sum_{k=0}^{\infty} e^{-s (2k)(2k+2)}.
\end{equation}
The first term on the right-hand side is understood as the contribution from the smooth part of $S^3/\mathbb{Z}_2$, which can be written as an integration over Seeley–De~Witt coefficients. The $\frac{1}{2}$ factor is due to the reduced volume by $\mathbb{Z}_2$ action. The second term can be understood as the contribution due to the orbifold singularity. To show that no additional logarithmic correction is induced from the orbifold singularity, it is sufficient to show that the contribution does not contain an $\mathcal{O}\left(s^0\right)$ piece in the small $s$ expansion. 

To see that, use the Euler–Maclaurin formula on the sum 
\begin{equation}
\begin{split}
\sum_{k=0}^\infty e^{-s (2k)(2k+2)}&=\frac{1}{2}+\int_0^{\infty}\mathrm{d}x\, e^{-s (2x)(2x+2)}+\sum_{k=2}^{\infty}\frac{B_k}{k!}\left(\frac{\mathrm{d}}{\mathrm{d}x}\right)^{k-1}e^{-s(2x)(2x+2)}|_{x=0}\\&=\frac{1}{2}+\frac{ e^s }{4}\sqrt{\frac{\pi}{s} }\left(1-\text{erf}\left( \sqrt{s}\right)\right)+\sum_{k=2}^{\infty}\frac{B_k}{k!}\left(\frac{\mathrm{d}}{\mathrm{d}x}\right)^{k-1}e^{-s(2x)(2x+2)}|_{x=0},
\end{split}
\end{equation}
where $B_k$ denotes the $k$th Bernoulli number. We note that no $\mathcal{O}\left(s^0\right)$ term is contained in the infinite sums on the right-hand side as the derivative leads to powers of $s.$ The first two terms combine to give an expansion 
\begin{equation}
    \frac{\sqrt{\pi}}{4}\frac{1}{s^{\frac{1}{2}}}+\frac{\sqrt{\pi}}{4}s^{\frac{1}{2}}+\mathcal{O}(s^{\frac{3}{2}}),
\end{equation}
where there is no $s^0$ term. Thus no additional logarithmic correction is induced due to the orbifold singularity for the scalar field.

One can also compute the heat kernel of other types of fields on $S^3/\mathbb{Z}_2$ as in \cite{David:2009xg}.

\section{Topological data}\label{app:hodge}

For the ES duality, the integer homology and cohomology groups
of $(S^3\times K3)/\mathbb{Z}_2$ are  useful. These are listed below.

\begin{eqnarray}
H_0((S^3\times K3)/\mathbb{Z}_2;\mathbb{Z})&=\mathbb{Z}, \quad
&H^0((S^3\times K3)/\mathbb{Z}_2;\mathbb{Z})=\mathbb{Z},\\
H_1((S^3\times K3)/\mathbb{Z}_2;\mathbb{Z})&=\mathbb{Z}_2, \quad
&H^1((S^3\times K3)/\mathbb{Z}_2;\mathbb{Z})=0,\\
H_2((S^3\times
K3)/\mathbb{Z}_2;\mathbb{Z})&=\mathbb{Z}^{10}+\mathbb{Z}_2,\quad
&H^2((S^3\times K3)/\mathbb{Z}_2;\mathbb{Z})=\mathbb{Z}^{10}+\mathbb{Z}_2,\\
H_3((S^3\times K3)/\mathbb{Z}_2;\mathbb{Z})&=\mathbb{Z}, \quad
&H^3((S^3\times K3)/\mathbb{Z}_2;\mathbb{Z})=\mathbb{Z}+\mathbb{Z}_2,\\
H_4((S^3\times
K3)/\mathbb{Z}_2;\mathbb{Z})&=\mathbb{Z}+\mathbb{Z}_2, \quad
&H^4((S^3\times K3)/\mathbb{Z}_2;\mathbb{Z})=\mathbb{Z},\\
H_5((S^3\times
K3)/\mathbb{Z}_2;\mathbb{Z})&=\mathbb{Z}^{10}+\mathbb{Z}_2, \quad
&H^5((S^3\times K3)/\mathbb{Z}_2;\mathbb{Z})=\mathbb{Z}^{10}+\mathbb{Z}_2,\\
H_6((S^3\times K3)/\mathbb{Z}_2;\mathbb{Z})&=0, \quad
&H^6((S^3\times K3)/\mathbb{Z}_2;\mathbb{Z})=\mathbb{Z}_2,\\
H_7((S^3\times K3)/\mathbb{Z}_2;\mathbb{Z})&=\mathbb{Z}, \quad
&H^7((S^3\times K3)/\mathbb{Z}_2;\mathbb{Z})=\mathbb{Z}.
\end{eqnarray}
\vspace{.02cm}

\bibliographystyle{JHEP}
\bibliography{ESduality}

\end{document}